\documentclass[12pt]{article}
\input{amssymb.sty}

\def\thefootnote{\fnsymbol{footnote}}
\catcode`\@=11
\def\numberbysection{\@addtoreset{equation}{section}
        \def\theequation{\thesection.\arabic{equation}}}
\numberbysection

\def\beq{\begin{equation}}
\def\eeq{\end{equation}}
\def\bea{\begin{eqnarray}}
\def\eea{\end{eqnarray}}

\def\half{{\textstyle {1\over 2}}}
\def\e{\varepsilon}
\def\s{\sigma}
\def\z{\zeta}
\def\Q{{\Bbb Q}}
\def\Z{{\Bbb Z}}
\def\N{{\Bbb N}}
\def\H{{\cal H}}
\def\R{{\cal R}}
\def\KT{{\rm KT}}

\renewcommand{\rho}{\varrho}
\begin{document}

\pagenumbering{alph}
\setcounter{page}{0}
\renewcommand{\thefootnote}{$^\dagger$}

\rightline{UCL--IPT--98--03}

\vskip 4cm
{\LARGE \centerline{Discrete symmetries of unitary minimal}
\centerline{conformal theories}}

\vskip 3cm

\centerline{\large P. Ruelle\footnote{Chercheur
Qualifi\'e FNRS} and O. Verhoeven}

\vskip 2truecm
\centerline{Institut de Physique Th\'eorique}
\centerline{Universit\'e Catholique de Louvain}
\centerline{B--1348 \hskip 0.5truecm Louvain-La-Neuve, Belgium}

\vskip 3truecm
\begin{abstract} 
\noindent
We classify the possible discrete (finite) symmetries of
two--dimensional critical models described by unitary minimal conformally
invariant theories. We find that all but six models have the group $Z_2$ as
maximal symmetry. Among the six exceptional theories, four have no symmetry at
all, while the other two are the familiar critical and tricritical 3--Potts
models, which both have an $S_3$ symmetry. These symmetries are the expected
ones, and coincide with the automorphism groups of the Dynkin diagrams of
simply--laced simple Lie algebras ADE. We note that extended chiral algebras,
when present, are almost never preserved in the frustrated sectors.
\end{abstract}

\renewcommand{\thefootnote}{\arabic{footnote}}
\setcounter{footnote}{0}
\newpage
\pagenumbering{arabic}
\baselineskip=16pt


\section{Introduction}

Critical models with infinite correlation lengths fall in universality classes
according to their long range behaviours. In the vicinity of the critical
point, the long range correlations, in their scaling limit, are those of
a continuous field theory, which becomes massless, conformally invariant at
the critical point.  

In two dimensions, conformal field theories have much more structure than in
higher dimensions. In order to expose their full content, one may put the
conformal field theory in various geometries, with various boundary conditions.
In particular, the consistent formulation of a conformal theory on the torus has
proved to be extremely constraining for the theory itself \cite{cardy1}.
Consistent means, among other things, that the periodic partition function must
take the same value on conformally equivalent tori: it must be modular
invariant.

It turns out that modular invariance is formidably restrictive, putting
very strong constraints on the field content of the theory. It is in fact so
restrictive that a classification program of all consistent conformal theories
has been initiated, following the seminal ideas of \cite{cardy1}. A first step
towards this vastly ambitious goal was taken in \cite{ade,kato}, where a
complete list was given of all possible modular invariant partition functions of
so--called minimal conformal theories. Remarkably, this list is structured in a
few infinite series that follow an ADE pattern.

For the unitary theories, it is now established \cite{lattice}
that the entries in the ADE list, restricted to the unitary cases, are
realized by actual lattice models \cite{pasquier,dilute}, namely the RSOS and
later the dilute RSOS models, all defined in terms of ADE Dynkin diagrams. It
means that for every modular invariant partition function of the list, there is
(at least) one statistical lattice model whose critical partition function is
the given item in the list. In some cases, different lattice models are known
which have their critical partition function equal to the same modular
invariant. 

None of these models has a continuous symmetry group, but most of them have
finite symmetries \cite{cardy2,vonG_R}. Well--known cases include
the Ising model, with a $Z_2$ symmetry, and the 3--Potts model, with an $S_3$
symmetry. Zuber \cite{zuber} was the first to investigate, in a systematic way,
the presence of a discrete symmetry from the knowledge of the critical modular
invariant ({\it i.e.} periodic) partition function. It is our purpose to pursue
his analysis, and indeed to determine the maximal symmetry group of all unitary
minimal conformal theories. The importance of knowing the symmetry is two--fold.

On one hand, it helps identify the conformal theory describing the critical
regime of a lattice model\footnote{Unfortunately almost all models we examine
have the same symmetry, so its knowledge does not help much here.} (or
vice versa). The symmetries we determine are those of continuous field
theories, which are the continuum limits of critical lattice models. They are
generally expected to be the symmetries already present on the lattice. Indeed
in the present case, the symmetries we find are realized in the discrete RSOS
models. Thus the symmetry of a given conformal theory is presumably shared by
all the lattice models in the universality class corresponding to that
conformal theory.

On the other hand, the question arises precisely as to what extent a modular
invariant partition function specifies a unique universality class (a conformal
theory). Indeed it only fixes the content of the periodic sector, and it is not
clear that this is saying anything about the non--periodic sectors that might
be consistently added. Apart from unconsistencies that may come from other
sources than modular covariance, we see three basic reasons why the univocity
of the association of a conformal field theory with a modular invariant
partition function may fail.

First, it is conceivable that a partition function be compatible with a
specific symmetry, but in more than one way. In this case, one would conclude 
that different realizations of the symmetry lead to inequivalent conformal
theories hence to different universality classes, since the field contents and
the charges are different. Our results show that this situation does not occur
(in the models that we have examined): when a symmetry is present, it is always
realized in exactly one way. 

A second situation is when the partition function is compatible with a
given symmetry, which however is not fully exploited. In other words, the
Hilbert spaces corresponding to twisted boundary conditions, that could, {\it
in principle}, be adjoined to the periodic sector to form a consistent
conformal theory, simply do not exist. After all, the periodic sector of any
model is self--consistent, and there is no way to say whether or not the
frustration operators are present in the lattice model. Indeed there is no 
reason as to why a model should be forced to use the maximal symmetry which is
available. 

The previous two situations somehow rely on the point of view that the
non--periodic sectors are organized by a symmetry, which is also the stand we
take in this article. That is to say, the non--periodic boundary conditions are
obtained from a group operation on the microscopic variables that leaves the
periodic Hamiltonian invariant. There might be a last possibility that the
different sectors are not related to the existence of a symmetry. 

Examples of models falling in the last two categories have been suggested in
\cite{agr}, where $c<1$ unitary models were obtained through suitable
projections of the spectrum of the XXZ Heisenberg chains. However no example 
was given of lattice models with the predicted spectra. 

Finally, we would like to stress the fact that the symmetry groups we determine
are computed with respect to a built--in chiral algebra, which is here the
conformal algebra. This is made manifest because our analysis relies on the
assumption that all Hilbert spaces decompose into Virasoro representations, so
that all partition functions are expressed in terms of conformal characters.
Another starting point, based for instance on an extended chiral
algebra, may lead to different results. Examples of such situations will be
encountered in Section 6, where block diagonal modular invariant partition
functions are considered.


\section{Statement of the problem and results}

The definition of a lattice model requires to specify the boundary conditions.
In a toroidal geometry, one identifies the opposite edges of the
rectangle (soon to become a parallelogram), and the boundary conditions say
how the spin configurations on these edges are related. The periodic boundary
conditions are the simplest ones: the configurations on opposite edges are
equal.

If the Hamiltonian (or the Boltzmann weights) for the periodic boundary
conditions are invariant under a finite symmetry group $G$, one may impose
twisted boundary conditions in which the spin variables on opposite edges are
related by group operations. On an $L \times M$ rectangular lattice, they are
\beq
\sigma_{L+1,j} = {}^{g}\sigma_{1,j}\,, \qquad 
\sigma_{i,M+1} = {}^{g'}\!\sigma_{i,1}\,, \qquad \hbox{ with $g,g' \in G$}.
\label{tbc}
\eeq
Equivalently, twisted boundary conditions can be implemented by the insertion,
in the periodic system, of frustration lines. 

For each boundary condition, one can compute a partition function
$Z_{g,g'}(L,M)$. Let us note that one should restrict to twisting
elements $g,g'$ that commute, because $\s_{L+1,M+1}$ $= {}^{gg'}\s_{1,1} =
{}^{g'g}\s_{1,1}$. Furthermore, if $G$ is non--Abelian (like in the 3--Potts
model), the symmetry of the model with the boundary conditions $(g,g')$ is
broken down\footnote{In that sense, an extended chiral algebra can also be
broken in the non--periodic sectors, down to the conformal algebra. Examples
include the critical 3--Potts model, where the $W_3$ algebra is broken by the
antiperiodic boundary conditions, see Section 6.4.} to a subgroup of
$G$, which is the centralizer of $\{g,g'\}$. The reason is that if the
Hamiltonian in the bulk has the invariance $H\{\s\}=H\{^g\s\}$ for all $g \in
G$, the boundary conditions (\ref{tbc}) may not be invariant under all of $G$.

In the continuum limit, the density of lattice points increases, and one
obtains a full rectangle, of sides $L$ and $iM$ in the complex plane. The
underlying field theory being scale invariant, the partition
functions can only depend on the ratio $\tau=iM/L$. Equivalently, one may
normalize $L$ to 1, and consider the rectangle with sides 1 and $\tau$, where
$\tau$ is purely imaginary and $-i\tau$ strictly positive. One can be more
general and take an arbitrary torus, whose shape is a parallelogram rather than
a rectangle. In the complex plane, this amounts to give the two independent
periods 1 and $\tau$, and the identifications $z \simeq  z+1$ and $z \simeq
z+\tau$. The modulus $\tau$ now belongs to the upper half plane, Im$\,\tau > 0$.
The boundary conditions then relate the field configurations on the opposite
sides of the parallelogram, 
\beq
\phi(z+1)={}^g\phi(z), \qquad \phi(z+\tau)={}^{g'}\!\phi(z),
\eeq
and define the frustrated partition functions $Z_{g,g'}(\tau)$.

In conformal field theory, these functions can be computed in the Hamiltonian
formalism ({\it i.e.} the transfer matrix formalism), with the following result
\cite{cardy1,cardy2,itz-zub}
\beq
Z_{g,g'}(\tau) = {\rm Tr}_{\H_g} \, \Big(q^{(L_0-c/24)} \, \overline
q^{(\overline L_0-c/24)}g'\Big), \qquad q=e^{2i\pi \tau}.
\eeq
In this expression, $\H_g$ is the Hilbert space of states that live on the
fixed time slices (lines of constant Im$\,z$ in the complex plane) in the
presence of boundary conditions twisted by $g$ in space. It encodes the
information about the boundary condition in space in the same way that a
row--to--row transfer matrix includes the boundary condition along a single
row. Thus $\H_g$ determines the state or field content of the theory in the
$g$--sector. The boundary condition in time is effected by the insertion of
$g'$. The operators $L_0$ and $\overline L_0$ are the generators of dilations in
the $z$ and $\overline z$ variables, combinations of which yield the Hamiltonian
and momentum operators. Finally the number $c$ is the central charge of the
conformal theory in question, namely the central term occurring in the Virasoro
algebra. 

Conformal symmetry implies that each space $\H_g$ is made up of representations
of a pair of Virasoro algebras Vir$_c \times {\rm Vir}_c$ with equal central
term $c$
\beq
\H_g = \bigoplus_{i,j} \; M_{ij}^{(g)} \; (\R_i \otimes \R_j),
\eeq
where the numbers $M_{i,j}^{(g)}$ are multiplicities, {\it i.e.} they are 
non--negative integers. The labels $i,j$ specify the inequivalent
representations of Vir$_c$. 

Assume now that the theory has the discrete symmetry group $G$. It means that
its energy--momentum tensor is left invariant by $G$, with the consequence that
the group action can only mix together equivalent representations occurring in
each $\H_g$. So if $g'$ has order $N$, and if $\R_{i} \otimes \R_j$ occurs in
$\H_g$ with multiplicity $M_{i,j}^{(g)}$, the action of $g'$ on them is by some matrix
of order $N$. That matrix can be diagonalized, and yields a diagonal
action by $N$--th roots of unity\footnote{In concrete models, the group of
symmetry has some definite action on the variables/primary fields, which leads
to a preferred basis. From the pure representation theory, any choice of basis
in the degenerate modules $(\R_i \otimes \R_j)_k$ is as good as any other.}.
Thus one can write, in the $g$--sector,
\beq
g'(\R_i \otimes \R_j)_k = \z_N^{Q(g;i,j;k;g')} \, (\R_i \otimes \R_j)_k\,, 
\qquad k=1,2,\ldots,M^{(g)}_{ij},
\eeq
with $\z_N = e^{2i\pi/N}$. The integer $Q(g;i,j;k;g')$, defined modulo $N$, can
be called a $g'$--charge (or a parity for an order 2 element). 

Using now the definition of the Virasoro characters $\chi_i(q) = {\rm Tr}_{\R_i}
\, q^{(L_0-c/24)}$, the decomposition of the spaces $\H_g$ and the action of the
group on their content, one obtains the following form for the frustrated
partition functions on a torus of modulus $\tau$
\beq
Z_{g,g'}(\tau) = \sum_{i,j} \; \Big[\sum_{k=1}^{M_{ij}^{(g)}} \;
\z_N^{Q(g;i,j;k;g')}\Big] \; \chi^*_{i}(q) \, \chi^{}_{j}(q),
\label{tpf}
\eeq
from which the charges of the various fields can be easily read off.
This formula is the first fundamental ingredient of our analysis.

From the conformal point of view, the description of tori in terms of complex
moduli $\tau$ is redundant. Indeed $\tau$ and $\tau + 1$ clearly specify the
same torus because the identifications $z \simeq z+1 \simeq z+\tau+1$ are
manifestly equivalent to $z \simeq z+1 \simeq z+\tau$. Moreover the moduli
$\tau$ and $-{1 \over \tau}$ determine conformally equivalent tori. This may be
less apparent but it follows from the conformal transformation $w=-{z \over
\tau}$, in terms of which the identifications become $w \simeq w-1 \simeq w
-{1 \over \tau}$. The corresponding two transformations, $T\;:\; \tau
\rightarrow \tau+1$ and $S\;:\; \tau \rightarrow -{1 \over \tau}$, generate the
modular group $PSL_2(\Z)$
\beq
PSL_2(\Z) = \Big\{\tau \rightarrow {a\tau + b \over c\tau + d} \;:\;
a,b,c,d \in \Z \; {\rm and} \; ad-bc=1\Big\},
\label{psl2}
\eeq
which can also be presented as $PSL_2(\Z) = \langle S,T \,|\, S^2 = (ST)^3 =1
\rangle$. Two tori are conformally equivalent if and only if their moduli are
related by a modular transformation of $PSL_2(\Z)$. 

The existence of modular transformations has immediate and dramatic consequences
on the frustrated partition functions \cite{cardy1,zuber}. Being partition
functions of conformally invariant theories, they should all be invariant under
modular transformations, except for a change in the boundary conditions. Under
$\tau'=\tau+1$, the field configurations are related by $\phi(z+\tau') =
{}^{gg'}\!\phi(z)$, so that the effective twist is $gg'$ along $\tau'$. The
other transformation, $\tau'=-{1 \over \tau}$, is implemented by the
conformal transformation $w=-{z \over \tau}$, which has the property of
exchanging the two periods: the identification $z \simeq z+1$ goes to
$w \simeq w+\tau'$, while $z \simeq z+\tau$ becomes $w \simeq w-1$. Thus it
follows that the two twists are in effect exchanged, with one of them inversed.
Putting all together, one finds the transformation law of the partition
functions under the modular group
\cite{zuber}:
\beq
Z_{g,g'}(\tau) = Z_{g,gg'}(\tau+1) = Z_{g'^{-1},g}({\textstyle -{1 \over
\tau}}) = Z_{g^ag'^c,g^bg'^d}({\textstyle {a\tau+b \over c\tau+d}}).
\label{tmod}
\eeq
This is the second fundamental ingredient. Let us note that the transformation
$\tau \rightarrow -{1 \over \tau}$ being an involution implies the identities
$Z_{g,g'}(\tau) = Z_{g^{-1},g'^{-1}}(\tau)$. Thus the torus partition functions
transform properly under $PSL_2(\Z)$, even in the presence of boundary
conditions. 

One sees that the partition functions are not individually invariant under the
full modular group but only under the subgroup of $PSL_2(\Z)$ which fixes
their boundary conditions \cite{zuber}. Depending on $g,g'$, the subgroups are
various congruence subgroups of level $N$ if the orders of the elements of $G$
are divisors of $N$. For a generic pair $g,g'$, the individual function
$Z_{g,g'}$ is invariant under the level $N$ principal congruence subgroup
$\Gamma(N)$ ($\big({a \atop c}\;{b \atop d}\big) = \pm \big({1 \atop 0}\;{0
\atop 1} \big) \bmod N$), while $Z_{g,1}$ and $Z_{1,g'}$ are respectively
invariant under $\hat\Gamma^0(N)$ ($\big({a \atop c}\; {b \atop d}\big) =
\pm \big({1 \atop c}\;{0  \atop 1}\big) \bmod N$) and $\hat\Gamma_0(N)$
($\big({a \atop c}\; {b \atop d}\big) = \pm \big({1 \atop 0}\;{b \atop 1}\big)
\bmod N$). Finally the partition function for periodic boundary conditions,
$Z_{1,1}(\tau)$, is fully modular invariant. Each of these subgroups has finite
index in the modular group; the precise action of the cosets on the set of
functions $Z_{g,g'}$ was discussed in \cite{zuber}. 

The main question that now arises is whether the general form (\ref{tpf}) of
the partition functions is consistent with the modular transformations
(\ref{tmod}). In a way, the answer was the main result of Cardy in
\cite{cardy1}: the two requirements are not compatible, unless the field
content of the various sectors, the possible symmetries and the charges of the
fields under that symmetry are chosen in a very specific way. This explains
why the requirement of modular invariance or covariance is so constraining and
powerful. Our purpose is to find all solutions to this problem for the
particular class of unitary minimal conformal theories.

Unitary minimal conformal theories are among the simplest examples of rational
theories (for general background, see \cite{dms}). They have a central charge
given by $c(m)=1-{6 \over m(m+1)}<1$, where $m \geq 3$ is an integer. For fixed
$m$, the Virasoro algebra with central term $c=c(m)$ has finitely many unitary
representations, labelled by pairs $(r,s)$ of integers satisfying $1
\leq s \leq r \leq m-1$. The characters of these representations are known
functions of $\tau$ \cite{rocha}, and their modular transformations can be
explicitly computed (see below). Therefore for these theories, the problem is
posed in concrete terms. 

A first, substantial step is to look at all possible periodic partition
functions. One looks for modular invariant functions, which are
sesquilinear forms in the Virasoro characters, of the type
\beq 
Z_{1,1} = \sum_{(r,s),(r',s')} \; [\chi_{(r,s)}(\tau)]^* \,
M^{(1)}_{(r,s),(r',s')} \, [\chi_{(r',s')}(\tau)],
\label{mipf}
\eeq
for non--negative integral coefficients $M^{(1)}_{(r,s),(r',s')}$. The
representation $\R_{(1,1)} \otimes \R_{(1,1)}$ is the only one to contain the
vacuum of the theory. Since the vacuum is certainly in the periodic sector
$\H_1$, its unicity requires that none of the other spaces $\H_g$, $g \neq
1$, contain it. One thus imposes $M^{(g)}_{(1,1),(1,1)}=\delta_{g,1}$, which
fixes the normalization of $Z_{1,1}$.

The classification of all modular invariant functions with these properties
was accomplished by Cappelli, Itzykson and Zuber \cite{ade}, and by Kato
\cite{kato}, and remains one of the most remarkable results in conformal
theory. The functions $Z_{1,1}$, labelled by pairs of simply--laced Lie
algebras, are listed in Table 1. By using the symmetry $\chi_{(r,s)} =
\chi_{(m-r,m+1-s)}$, one has for convenience extended the range of $(r,s)$ to
$\{1,\ldots,m-1\} \times \{1,\ldots,m\}$, which covers twice the original set
(called the Kac table). 

\begin{table}
\renewcommand{\arraystretch}{1.8}
\begin{center}
\begin{tabular}[t]{|c|l|c|}
\hline
$m$ & Periodic partition function $Z_{1,1}$ & Name \\
\hline \hline
$m \geq 3$ & $\half \,{\displaystyle {\sum_{r=1}^{m-1} \, \sum_{s=1}^m \;
|\chi_{(r,s)}|^2}}$ & $(A_{m-1},A_m)$ \\
\hline
\mbox{{\footnotesize $m=4\ell+1$}} & $\half \,{\displaystyle {\sum_{r=1}^{m-1}
\Big\{\sum_{s=1 \atop {\rm odd}}^{2\ell -1} |\chi_{(r,s)}+\chi_{(r,m+1-s)}|^2 
+2 |\chi_{(r,2\ell + 1)}|^2 \Big\}}}$ &
\mbox{{\footnotesize $(A_{m-1},D_{2\ell+2})$}} \\
\hline
\mbox{{\footnotesize $m=4\ell+2$}} & $\half \,{\displaystyle {\sum_{s=1}^{m}
\Big\{\sum_{r=1 \atop {\rm odd}}^{2\ell -1} |\chi_{(r,s)}+\chi_{(m-r,s)}|^2 
+2 |\chi_{(2\ell + 1,s)}|^2 \Big\}}}$ & \mbox{{\footnotesize
$(D_{2\ell+2},A_{m})$}} \\
\hline
\mbox{{\footnotesize $m=4\ell+3$}} & $\half \,{\displaystyle {\sum_{r=1}^{m-1}
\Big\{\sum_{s=1 \atop {\rm odd}}^m |\chi_{(r,s)}|^2 + |\chi_{(r,2\ell+2)}|^2
+ \sum_{{\rm even}\ s=2 \atop s \neq 2\ell+2}^m
\chi_{(r,s)}^*\,\chi_{(r,m+1-s)}
\Big\}}}$ & \mbox{{\footnotesize $(A_{m-1},D_{2\ell+3})$}} \\
\hline
\mbox{{\footnotesize $m=4\ell+4$}} & $\half \,{\displaystyle {\sum_{s=1}^{m}
\Big\{\sum_{r=1 \atop {\rm odd}}^{m-1} |\chi_{(r,s)}|^2 +
|\chi_{(2\ell+2,s)}|^2 + 
\sum_{{\rm even}\ r=2 \atop r \neq 2\ell+2}^{m-1}
\chi_{(r,s)}^*\,\chi_{(m-r,s)} \Big\}}}$ & \mbox{{\footnotesize
$(D_{2\ell+3},A_m)$}} \\
\hline
$m=11$ & $\half \,{\displaystyle {\sum_{r=1}^{10} \;
|\chi_{(r,1)}+\chi_{(r,7)}|^2 + |\chi_{(r,4)}+\chi_{(r,8)}|^2 +
|\chi_{(r,5)}+\chi_{(r,11)}|^2 }}$ & $(A_{10},E_6)$ \\
\hline
$m=12$ & $\half \,{\displaystyle {\sum_{s=1}^{12} \;
|\chi_{(1,s)}+\chi_{(7,s)}|^2 + |\chi_{(4,s)}+\chi_{(8,s)}|^2 +
|\chi_{(5,s)}+\chi_{(11,s)}|^2}}$ & $(E_6,A_{12})$ \\
\hline
$m=17$ & $\half \,{\displaystyle {\sum_{r=1}^{16} \;
|\chi_{(r,1)}+\chi_{(r,17)}|^2 + |\chi_{(r,5)}+\chi_{(r,13)}|^2 +
|\chi_{(r,7)}+\chi_{(r,11)}|^2}}$ &  \\
 & $+ |\chi_{(r,9)}|^2 + [\chi_{(r,3)}+\chi_{(r,15)}]^*\,\chi_{(r,9)} + 
\chi_{(r,9)}^*\,[\chi_{(r,3)}+\chi_{(r,15)}]$ & $(A_{16},E_7)$\\
\hline
$m=18$ & $\half \,{\displaystyle {\sum_{s=1}^{18} \;
|\chi_{(1,s)}+\chi_{(17,s)}|^2 + |\chi_{(5,s)}+\chi_{(13,s)}|^2 +
|\chi_{(7,s)}+\chi_{(11,s)}|^2}}$ &  \\
 & $+ |\chi_{(9,s)}|^2 + [\chi_{(3,s)}+\chi_{(15,s)}]^*\,\chi_{(9,s)} + 
\chi_{(9,s)}^*\,[\chi_{(3,s)}+\chi_{(15,s)}]$ & $(E_7,A_{18})$\\
\hline
$m=29$ & \mbox{{\footnotesize $\half \,{\displaystyle {\sum_{r=1}^{28} \;
|\chi_{(r,1)}\!+\!\chi_{(r,11)}\!+\!\chi_{(r,19)}\!+\!\chi_{(r,29)}|^2 \!+\! 
|\chi_{(r,7)}\!+\!\chi_{(r,13)}\!+\!\chi_{(r,17)}\!+\!\chi_{(r,23)}|^2}}$}} &
$(A_{28},E_8)$\\
\hline
$m=30$ & \mbox{{\footnotesize $\half \,{\displaystyle {\sum_{s=1}^{30} \;
|\chi_{(1,s)}\!+\!\chi_{(11,s)}\!+\!\chi_{(19,s)}\!+\!\chi_{(29,s)}|^2 \!+\! 
|\chi_{(7,s)}\!+\!\chi_{(13,s)}\!+\!\chi_{(17,s)}\!+\!\chi_{(23,s)}|^2}}$}} &
$(E_8,A_{30})$\\
\hline
\end{tabular}
\end{center}
\begin{center}
\begin{minipage}{14cm}
\vskip 0.2truecm
\baselineskip 13pt
{\small {\it Table 1.} Complete list of modular invariant partition functions
for unitary minimal conformal theories. The integer $m$ is related to the
central charge by $c=1-{6 \over m(m+1)}$.}
\end{minipage}
\end{center}
\end{table}

\renewcommand{\arraystretch}{1}
One can see from the table that, for fixed $m$, there is only a very limited
number of possible periodic partition functions, most often only two. Those of
the series $(A_{m-1},A_m)$ have been identified as describing the multicritical
points in RSOS models \cite{abf_h}. The 3--Potts model corresponds to the
$(A_4,D_4)$ theory \cite{dotsenko,cardy2,vonG_R}, and the tricritical 3--Potts
model to $(D_4,A_6)$. All of them are partition functions of dilute RSOS models
\cite{pasquier,dilute,lattice}.

The list of modular invariant partition functions is our starting point.
For each such function $Z_{1,1}$, we want to find all possible sets of functions
$Z_{g,g'}$ which satisfy the above two conditions: they must form a
closed set under the action of the modular group, and they must be of the form
(\ref{tpf}). The known function $Z_{1,1}$ fixes the form of all $Z_{1,g'}$,
which themselves yield other functions $Z_{g,g'}$ by modular transformations.
All these other functions must have the required properties. 

We have mentionned above that such sets are consistent only for group elements
$g,g'$ which commute, and which therefore generate an Abelian group. 
In practice one first looks for finite cyclic symmetry groups. The
existence of a cyclic $Z_N$ symmetry (of $N^2$ functions $Z_{g^i,g^j}$, with
$g$ the generator) implies that $G$ contains a cyclic subgroup. Subtleties can 
arise when several cyclic symmetries are compatible with a given partition
function. Suppose for example that a modular invariant function $Z_{1,1}$ is
compatible with two symmetries $Z_N$ and $Z_{N'}$. Namely, we have two sets 
\beq
\{Z_{g^i,g^j} \;:\; 1 \leq i,j \leq N\} \,, \quad  \hbox{and} \quad
\{Z_{g'^k,g'^l} \;:\; 1 \leq k,l \leq N'\},
\eeq
with the {\it same} $Z_{1,1}$. If one could find a consistent set of functions
$Z_{g^ig'^k,g^jg'^l}$, one would conclude that the two symmetries are
compatible --the two types of charges are simultaneously assignable--, and that
in fact the composite symmetry $Z_N \times Z_{N'}$ ($=Z_{NN'}$ if $N$ and $N'$
are coprime integers) is present. If not, the two cyclic groups are not
commuting subgroups. This means either that $G$ itself is a non--commutative
group though containing two cyclic subgroups, or else that there exist two
genuinely different models with the same periodic partition function. This
second alternative would correspond to a situation where the various Hilbert
spaces $\{\H_{g^i}\}$ and $\{\H_{g'^k}\}$ cannot be accomodated within a single
model. In principle, this question is decidable by looking at the periodic
sector, which fixes what the maximal symmetry is. One would try to find a
non--commutative action on the representations of the periodic sector, which
yields consistent partition functions and sensible modular transformations. That
action would be a representation of $G$. We will illustrate this in the only
two models where this situation occurs, see Section 6.4.

We have carried out this program for all unitary minimal models, with
the following results. {\it Every model in Table 1 has a unique, maximal
$Z_2$ symmetry, except six cases. The models $(A_4,D_4)$ (the critical 3--Potts
model) and $(D_4,A_6)$ (the tricritical 3--Potts model) have unique $Z_2$ and
$Z_3$ symmetries, which are not commuting and which
combine to an $S_3$ symmetry. The last four models in the Table, related to
$E_7$ and $E_8$, have no symmetry. Of all partition functions that can
be interpreted in terms of an extended chiral algebra (block diagonal form),
only the two with an $E_6$ label have their extended algebra that survives all
possible twisted (antiperiodic) boundary conditions. The extended algebra of
the critical and tricritical 3--Potts models is preserved only by the $Z_3$
twisted boundary conditions.} 

All explicit partition functions are given in the text. We have no explanation
for the peculiarity of the two $E_6$ models, regarding the full compatibility
of the extended algebra and the discrete symmetry group.

These results are expected. As mentionned before, the functions $Z_{1,1}$ we
start from are critical partition functions of dilute RSOS models. In their
very formulation, these models use data of simple Lie algebras of the type ADE,
and inherit the symmetry of their Dynkin diagrams. In other words, for
fixed ${\cal G}=A,D$ or $E$, the models $(A,{\cal G})$ and $({\cal G},A)$ do
have a symmetry group equal to the automorphism group of the Dynkin diagram of
${\cal G}$, always equal to $Z_2$, except for $D_4, E_7$ and $E_8$. The diagram
of $D_4$ has a symmetry $S_3$, and the $E_7,E_8$ have no symmetry. From that
point of view, our result is the statement that these known symmetries are
neither broken nor enhanced by the continuum limit. 

The rest of the paper is devoted to the proof of this statement.
We will not follow the approach of \cite{zuber}, which consists in looking
for submodular partition functions. Instead we make a strong use of
Galois theory, which we believe allows for a simpler analysis in this
context. The relevance of Galois theory stems from the algebraic nature of the 
representation of the modular group acting on the characters and of the
coefficients of the frustrated functions $Z_{g,g'}$. The importance in
conformal field theory of Galois techniques has been recognized in recent years
\cite{cg}. Section 3 recalls some general and known facts about these aspects,
while Section 4 contains certain results which are specific to the present
context. The subsequent sections contain the detailed analysis for the
different series in Table 1. 


\section{Modular transformations and Galois theory}

Unitary minimal conformal theories have central charges forming the
infinite sequence $c=1-{6 \over m(m+1)}$ for integer $m \geq 3$. For fixed $m$,
the Virasoro algebra has but a finite number of inequivalent representations,
labelled by pairs of integers $(r,s)$ in the Kac table
\beq
\KT = \{(r,s) \in \N^2 \;:\; 1 \leq s \leq r \leq m-1\}.
\eeq
The characters of these representations are denoted by $\chi_{(r,s)}(\tau)$. 

They depend on a complex variable $\tau$ lying in the upper half plane, on
which the modular group $PSL_2(\Z)$ acts. A remarkable feature of the
characters, valid much beyond the present context (see \cite{kacwaki}), is that
they transform linearly under the modular group. The (unitary) representation
they carry can be given in terms of the generators $S$ and $T$. Explicitly, one
finds 
\beq
\chi_{i}({\textstyle {-1 \over \tau}}) = \sum_{j \in \KT} \;
S_{i,j} \, \chi_{j}(\tau), \qquad 
\chi_{i}(\tau + 1) = \sum_{j \in \KT} \; T_{i,j} \,
\chi_{j}(\tau),
\eeq
where the matrices $S$ and $T$ are given by
\bea
&& S_{(r,s),(r',s')} = {\textstyle \sqrt{8 \over m(m+1)}} \; (-1)^{(r+s)(r'+s')}
\; \sin{\pi rr' \over m} \, \sin{\pi ss' \over m+1}, \label{sel} \\
\noalign{\medskip}
&& T_{(r,s),(r',s')} = e^{2i\pi (h_{r,s}-{c \over 24})} \, \delta_{r,r'} \,
\delta_{s,s'}\,, \qquad  h_{r,s} = {\textstyle {[(m+1)r-ms]^2-1 \over 4m(m+1)}}.
\label{tel}
\eea

One may check that $S$ and $T$ are unitary and symmetric, and that they satisfy
the defining relations of the modular group, $S^2=(ST)^3=1$. Moreover, one may
check that $S$ has the following symmetries, of which we will make a repeated
use in subsequent sections,
\beq
S_{(m-r,s),(r',s')} = S_{(r,m+1-s),(r',s')} = (-1)^{(m+1)r'+ms'+1} \,
S_{(r,s),(r',s')}.
\label{symm}
\eeq
However their most distinctive feature is that their entries are all algebraic
numbers (satisfy polynomial equations with rational coefficients). This also is
not specific to the present situation, but is known to hold in any rational
conformal theory \cite{deBG}, and prompts the use of Galois theory. The
following Galois properties of the modular matrices $S$ and $T$ have been
exposed in \cite{cg}.

The entries of $S$ and $T$ belong to some cyclotomic extension of the
rationals $\Q(\z_n)$, that is, they are linear combinations of $n$--th roots of
unity with rational coefficients. The Galois group of $\Q(\z_n)$ (over $\Q$),
namely the group of automorphisms of $\Q(\z_n)$ which fix $\Q$ pointwise, is
isomorphic to the group $\Z^*_n$ of invertible integers modulo $n$ (the
integers between 1 and $n$ which are coprime with $n$). This group has order
$\varphi(n)$, the Euler function. For $h \in \Z^*_n$, the Galois
transformation $\s_h$ acts on an $n$--th root of unity by raising it to
the $h$--th power: $\s_h(\z_n) = \z_n^h$. This action extends to $\Q(\z_n)$ by
linearity.  

It follows that cyclotomic Galois groups act on the coefficients of
$S$ and $T$. Here we will be more concerned with the action on $S$. The
formula (\ref{sel}) implies that the coefficients of $S$ lie in the extension
$\Q(\z_{2m(m+1)})$. It is a general fact that the action on $S$ of the
Galois group induces a permutation of the Kac table by the formula \cite{cg}
\beq
\s_h(S_{(r,s),(r',s')}) = \e_h(r,s) \, S_{\s_h(r,s),(r',s')} = 
\e_h(r',s') \, S_{(r,s),\s_h(r',s')},
\label{gal_s}
\eeq
where the factors $\e_h$ are equal to $\pm 1$. The formula (\ref{sel}) allows
to determine explicitly the permutation $\s_h(r,s)$ and the signs $\e_h(r,s)$.

One first computes the residues $\langle hr \rangle_{2m}$ and $\langle hs
\rangle_{2(m+1)}$ of $hr$ modulo $2m$ and of $hs$ modulo $2(m+1)$ (we choose
the residues $\langle x \rangle_y$ between 0 and $y-1$). One then forms the
unique pair of integers $(\tilde r_h,\tilde s_h)$ in the rectangle $[1,m]
\times [1,m+1]$ by changing, if necessary, $\langle hr \rangle_{2m}$ by
$2m-\langle hr \rangle_{2m}$, and similarly $\langle hs \rangle_{2(m+1)}$ by
$2(m+1)-\langle hs \rangle_{2(m+1)}$. Up to a global sign related to the Galois
action on the radical $\sqrt{8/m(m+1)}$, the sign $\e_h(r,s)$ can be determined,
and is simply equal to $(-1)$ to the number of changes needed to go from
$(\langle hr \rangle_{2m},\langle hs \rangle_{2(m+1)})$ to
$(\tilde r_h,\tilde s_h)$. The pair $(\tilde r_h,\tilde s_h)$ is not in the Kac
table yet, but if not, then $(m-\tilde r_h,m+1-\tilde s_h)$ is in the Kac
table. 

Putting everything together, the Galois properties of $S$ are given by
(\ref{gal_s}) with
\bea
&& \s_h(r,s) =  \KT \cap \Big\{(\tilde r_h,\tilde s_h),(m-\tilde r_h,m+1-\tilde
s_h)\Big\},\\ 
\noalign{\medskip}
&& \e_h(r,s) = \eta_h(m) \, {\rm sign}\,(\sin{\textstyle {\pi hr \over
m}}\,\sin{\textstyle {\pi hs \over m+1}}),
\eea
where $\eta_h(m) = \s_h(\sqrt{8 \over m(m+1)})/\sqrt{8 \over m(m+1)}$ is a sign
which is independent of $r,s$.

The transformation law of the matrix $S$ under the Galois group has direct and
far--reaching consequences for the modular problem reviewed in the previous
section. Suppose we look for a modular invariant partition function of the type
(\ref{mipf}),
\beq 
Z_{1,1}(\tau) = \sum_{(r,s),(r',s')} \; [\chi_{(r,s)}(\tau)]^* \,
M^{(1)}_{(r,s),(r',s')} \, [\chi_{(r',s')}(\tau)], \qquad 
M^{(1)}_{(r,s),(r',s')} \in \N.
\eeq
Since it is invariant under $S$, one obtains, by using the
modular transformations of the characters, that the matrix $M^{(1)}$ must
satisfy $M^{(1)} = S^\dagger M^{(1)} S$, or $S M^{(1)} = M^{(1)} S$ because $S$
is unitary. Letting the Galois group act on this equation, a few
lines calculation shows that selection rules follow \cite{cg}
\beq
\hbox{$\e_h(r,s) \neq \e_h(r',s')$ for some $h$} \quad
\Longrightarrow \quad M^{(1)}_{(r,s),(r',s')} = 0.
\label{parule}
\eeq
The signs $\e_h(r,s)$ are generally referred to as (Galois) parities in the
literature, and the above set of conditions as the ``parity rule''. It gives
extremely strong constraints on $M^{(1)}$, because two pairs $(r,s)$ and
$(r',s')$ have only rarely the same parities for all $h$, implying that a huge
number of coefficients in $M^{(1)}$ are to vanish. Thus the matrices $M^{(1)}$
specifying modular invariant partition functions are usually sparse matrices. 

The next section collects the main implications that these algebraic
properties have for the existence of discrete symmetries in unitary minimal
models. 


\section{Galois constraints on frustrated systems}

This section contains four consequences of Galois theory which are
particularly relevant for our problem. Denoted R1 to R4, they are the most
fundamental arguments that will be used throughout the rest of the paper. We
collect them here to emphasize the fact that they are independent of the 
details of the periodic partition functions of Table 1. Some of them are based
on general Galois features only. 

Let us recall that we are primarily interested in cyclic groups of symmetry
$Z_N$. Let $g$ be a generator of $Z_N$. In order to simplify a bit the
notations, we define the matrices $M^{(i,j)}$ by
\beq 
Z_{g^i,g^j}(\tau) = \sum_{(r,s),(r',s')} \; [\chi_{(r,s)}(\tau)]^* \,
M^{(i,j)}_{(r,s),(r',s')} \, [\chi_{(r',s')}(\tau)].
\eeq
The periodic function $Z_{1,1}$ is taken to be one of the functions in Table 1,
and corresponds to a matrix $M^{(0,0)}$ with non--negative integral entries. The
matrix $M^{(0,1)}$ associated with the frustrated function $Z_{1,g}$ has the
same zero pattern as $M^{(0,0)}$, and is simply obtained from it by replacing
the non--zero coefficients by sums of $N$--th roots of unity:
\beq
M^{(0,0)}_{(r,s),(r',s')} = n \qquad \Longrightarrow \qquad 
M^{(0,1)}_{(r,s),(r',s')} = \z_N^{Q_1(r,s;r',s')} + \ldots +
\z_N^{Q_n(r,s;r',s')}.
\label{m01}
\eeq
Similarly, the matrix $M^{(0,i)}$ is obtained from $M^{(0,1)}$ by replacing all
roots of unity $\z_N$ by $\z_N^i$, effectively multiplying all charges $Q$ by
$i$. The matrices $M^{(i,0)}$ are related to $M^{(0,i)}$ by the modular
transformation $S$, according to (\ref{tmod}),
\beq
M^{(i,0)} = S^\dagger \, M^{(0,i)} \, S,
\eeq
and must also have non--negative integral coefficients. Finally, the matrices
$M^{(i,ki)}$ are related to $M^{(i,0)}$ by a power of the modular 
transformation $T$,
\beq
M^{(i,ki)} = T^k \, M^{(i,0)} \, T^{\dagger}{^k}.
\label{phases}
\eeq

The first statement R1 is easy to prove and provides a simple
although general means to determine the symmetry compatible with a given
model. In some cases, a crude look at the $S$ matrix allows to conclude. 

We start from the relation $M^{(1,0)} = S^\dagger M^{(0,1)} S$. The matrix on
the right--hand side is made up of numbers in the cyclotomic extension $\Bbb F$
containing the coefficients of $S$ and the $N$--th roots of unity, but, being
equal to $M^{(1,0)}$, is integer--valued. Therefore it is left invariant by the
Galois group of $\Bbb F$. Take any element $\s$ of the Galois group 
that fixes the matrix $S^\dagger \otimes S$. Since it also fixes $S^\dagger
M^{(0,1)}S$, the invertibility of $S$ implies that $\s$ actually fixes
$M^{(0,1)}$, that is, the matrix elements of $M^{(0,1)}$ must belong to the
field containing $S^\dagger \otimes S$. 

For the models we consider here, this simple result has an easy corollary. The
matrix $S$, given in (\ref{sel}), is real, and is therefore invariant under the
specific Galois transformation corresponding to complex conjugation. Hence
the matrix elements of $M^{(0,1)}$, given from  (\ref{m01}) by sums of $N$--th
roots of unity, must be real. The same is true of all $M^{(0,i)}$.

In case the matrix $M^{(0,0)}$ has all its coefficients in $\{0,1\}$, each
non--zero entry of $M^{(0,1)}$ is equal to a single $N$--th root of unity, which
must be real, so that $N$ can only be equal to 1 (no symmetry) or 2. It turns
out that all functions in Table 1, except those of the series
$(A_{m-1},D_{2\ell+2})$ and $(D_{2\ell+2,A_m})$, have their coefficients in
$\{0,1\}$. Thus, our first result says that 
\bea
\fbox{R1} && \hbox{\it If the modular matrix $S$ is real, a modular invariant
partition function having} \nonumber \\
\noalign{\vspace{-1mm}}
&& \hbox{\it all its coefficients in $\{0,1\}$ is compatible with a
cyclic symmetry $Z_2$ only.} \nonumber \\
\noalign{\vspace{-1mm}}
&& \hbox{\it As corollary, the maximal cyclic symmetry of minimal unitary 
models is $Z_2$,} \nonumber \\
\noalign{\vspace{-1mm}}
&& \hbox{\it except for the models in the complementary series for
$m=4\ell+1$ or $4\ell+2$.} \nonumber 
\eea
Let us stress that this statement is not equivalent to saying that $Z_2$ is the
maximal symmetry, since the $Z_2$ group could be realized in more than one
way, leaving the possibility for a power $(Z_2)^k$. 

\medskip
The second result is a strengthened version of the previous one, and relies
entirely on the Galois transformation of the matrix $S$, equation
(\ref{gal_s}). Let us emphasize again that this transformation is
completely general for a rational conformal theory. 

Like before, we start from $M^{(i,0)} = S^\dagger M^{(0,i)} S$, and write out
the transformations of the various matrices. For $\s$ any element of the Galois
group of $\Q(S_{ij})$, one has\footnote{From the result R1, $M^{(0,i)}$ is in
$\Q(S_{ij})$, so that its Galois group acts properly on it. As a consequence,
one obtains $\zeta_N \in \Q(S_{ij}) = \Q(\zeta_{2m(m+1)})$, that is, $N$
divides $2m(m+1)$ \cite{zuber}.} (summations over repeated indices)
\beq
(S^\dagger M^{(0,i)} S)_{kl} = S^\dagger_{k,\s(a)} \, \e_\s(a) \, (\s
M^{(0,i)})_{ab} \, \e_\s(b) \, S_{\s(b),l} = S^\dagger_{k,\s(a)} \, 
(\s M^{(0,i)})_{ab} \, S_{\s(b),l}
\eeq
where the last step follows from the parity rule: $M^{(0,i)}_{ab}$ is
non--zero if and only if $M^{(0,0)}_{ab}$ is non--zero, which requires
$\e_\s(a)\e_\s(b) = +1$ for all $\s$, see (\ref{parule}). Comparing the far
left-- and far right--hand sides leads to the conclusion that $\s$ acts on
$M^{(0,i)}$ by permutation of the indices: $(\s M^{(0,i)})_{ab} =
(M^{(0,i)})_{\s(a),\s(b)}$.

On the other hand, if one lets $\s$ act on the indices $k$ and $l$, one
obtains
\beq
M^{(i,0)}_{kl} = (S^\dagger M^{(0,i)} S)_{kl} = \s(S^\dagger M^{(0,i)}
S)_{kl} = \e_\s(k) \, \e_\s(l) \, (S^\dagger \s M^{(0,i)} S)_{\s(k),\s(l)}.
\eeq
Since $M^{(0,i)}$ contains $N$--th roots of unity, $\s$ acts on it by
substituting $\z_N^h$ for $\z^{}_N$ for some $h$, so in effect, $\s M^{(0,i)}
=  M^{(0,hi)}$. The previous equation then implies 
\beq
M^{(i,0)}_{kl} = \e_\s(k) \, \e_\s(l) \, M^{(hi,0)}_{\s(k),\s(l)}, \qquad 
\hbox{if $\s(\z^{}_N)=\z_N^h$}.
\eeq
This constraint is a generalization of the parity rule, to which it reduces
when $i=0$. Like in that case, it implies severe selection rules on the
coefficients of all matrices $M^{(i,0)}$.
\bea
\fbox{R2} && \hbox{\it The matrices $M^{(i,0)}$, specifying the field content
of the frustrated Hilbert spaces,} \nonumber \\
\noalign{\vspace{-0.5mm}}
&& \hbox{\it must all satisfy the parity rule: $M^{(i,0)}_{kl}=0$ unless
$\e_\s(k)\,\e_\s(l)=+1$ for all Galois} \nonumber \\ 
\noalign{\vspace{-0.5mm}}
&& \hbox{\it transformations. The matrices $M^{(0,i)}$, giving the charges of
the periodic sector,} \nonumber \\
&& \hbox{\it have Galois transformations $(\s M^{(0,i)})_{kl} =
(M^{(0,i)})_{\s(k),\s(l)}$.} \nonumber
\eea

\medskip
The third result we want to mention is by far the strongest: it solves
the parity rule, by saying exactly which matrix elements $M^{(i,0)}_{kl}$ may
be non--zero. The theorem we quote below is not properly new and has been
proved in \cite{aoki} (see \cite{gannon} for an elementary but clever proof
in the case $n$ odd). If the first two results can rightly be called
elementary, this one is not. Its use allows a straightforward proof of the
ADE classifications of conformal and $su(2)$ affine modular invariant partition
functions.

The basic problem is to determine all pairs $(r,s),(r',s')$ which have equal
Galois parities $\e_h(r,s)=\e_h(r',s')$ for all Galois transformations $\s_h$,
$h \in \Z^*_{2m(m+1)}$. A closed expression for (the $r,s$--dependent part
of) the parities has been given in Section 3,
\beq
\e_h(r,s) = {\rm sign}\,(\sin{\textstyle {\pi hr \over
m}}\,\sin{\textstyle {\pi hs \over m+1}}).
\eeq

Since $m$ and $m+1$ are coprime integers and since one of them is odd, one has a
canonical isomorphism $\Z^*_{2m(m+1)} = \Z^*_{2m} \times \Z^*_{2(m+1)}$, which
induces a similar splitting of the parity
\beq
\e_h(r,s) = {\rm sign}\,(\sin{\textstyle {\pi h_1r \over
m}})\,{\rm sign}\,(\sin{\textstyle {\pi h_2s \over m+1}}), \qquad 
h_1 \in \Z^*_{2m}, \; h_2 \in \Z^*_{2(m+1)}.
\eeq
Thus the problem effectively factorizes into two identical pieces. If one
defines the functions $\e_n(x) = {\rm sign}\,(\sin{\pi x \over n})$ for $x$ not
divisible by $n$, then 
\beq
\e_h(r,s)=\e_h(r',s') \quad \forall h \in \Z^*_{2m(m+1)} \qquad
\Longleftrightarrow \qquad 
\cases{\e_m(hr) = \e_m(hr') & $\forall h \in \Z^*_{2m}$, \cr
\e_{m+1}(hs) = \e_{m+1}(hs') & $\forall h \in \Z^*_{2(m+1)}$.}
\label{conf_su2}
\eeq
The function $\e_n(x)$ is actually the parity occurring in the Galois
transformation of the matrix $S$ relative to the characters of affine $su(2)$
algebras. As in the present situation, the affine $su(2)$ parities often appear
as constitutive blocks for the parities relative to the characters of other
algebras \cite{art}. 

Equation (\ref{conf_su2}) relates the conformal parity rule with the affine
$su(2)$ parity rule, the solution to the latter yielding the solution to the
former. It turns out that the $su(2)$ parity rule arises in mathematical
questions related to complex Fermat surfaces. (The same is true for the affine
$su(3)$ parity rule. See \cite{bcir} for a more precise description of the
connection between parity rules and problems in the geometry of Riemann
surfaces.) The results obtained by Aoki \cite{aoki} in that context actually
solve the $su(2)$ problem\footnote{The problem is not formulated in the same
way, but Theorem 2.7 of \cite{aoki} answers the question, except for finitely
many values of $n$. One can also set up a proof based on earlier results by
Koblitz and Rohrlich \cite{kr}.}. This will be our R3.
\bea
\fbox{R3} && \hbox{\it Suppose that $x$ and $y$, two integers between $1$ and
$n-1$, satisfy $\e_n(hx) = \e_n(hy)$} \nonumber \\ 
&& \hbox{\it for all $h \in \Z^*_{2n}$. Then, for GCD$(n,x,y)=1$, $x$ and $y$
must be related as follows:} \nonumber \\
&& \hbox{\it \hspace{1cm} (i) $\; y = x$ or $y = n - x$;} \nonumber \\
&& \hbox{\it \hspace{1cm} (ii) $\; n=6\;:\; x,y \in \{1,3,5\}$;} \nonumber \\
&& \hbox{\hspace{1.9cm} $n=10,12 \;:\; x,y \in \Z^*_n\,;$} \nonumber \\
&& \hbox{\it \hspace{1.9cm} $n=30 \;:\; x,y \in \{1,11,19,29\}$ or $x,y \in 
\{7,13,17,23\}.$} \nonumber\\
&& \hbox{\it All other solutions follow from these by multiplying $x,y$ and $n$
by a common} \nonumber \\
&& \hbox{\it integer.} \nonumber
\eea

\medskip
Our last general result concerns the use of R3 to further constrain the
possible symmetries, regardless of what the periodic partition function
is. In this sense, it strengthens R1.

The basic idea is to use the equation (\ref{phases}) in conjunction with R3.
Equation (\ref{phases}) says that $M^{(i,ki)}=T^k M^{(i,0)} T{^\dagger}{^k}$.
Because $T$ is diagonal, the entries of $M^{(i,ki)}$ and $M^{(i,0)}$ are
simultaneously zero or non--zero. The only difference between the non--zero
coefficients of the two matrices are some roots of unity, specifying the type
of symmetry and the charges of the sector $\H_{g^i}$ of the theory. The
conjugation by $T$ is what precisely produces these roots of unity since
\beq
M^{(i,ki)}_{(r,s),(r',s')} = e^{2i\pi k(h_{r,s} - h_{r',s'})} \,
M^{(i,0)}_{(r,s),(r',s')}.
\label{ttransf}
\eeq
In order to see what roots of unity can appear, it remains to compute the
quantities $h_{r,s} - h_{r',s'} \bmod 1$ for those pairs $(r,s),(r',s')$
for which $M^{(i,0)}_{(r,s),(r',s')}$ is non--zero. From R2, these pairs have 
to satisfy the parity rule, and are listed in R3.

In conformal models, the formula (\ref{tel}) yields
\beq
h_{r,s} - h_{r',s'} = {(r^2-r'^2)(m+1)^2 + (s^2-s'^2)m^2 \over 4m(m+1)} 
+ {rs + r's' \over 2} \; \bmod 1.
\eeq
Let us assume that $m$ is even, the analysis being the same for $m$ odd.

For $m$ even, we learn from R3 that $s'=s$ or $m+1-s$, and that the list
of pairs $r,r'$ satisfying the parity rule is the same as the list $r,m-r'$
that satisfy it. Since $h_{r',s'} = h_{m-r',m+1-s'}$, we can, without loss of
generality, assume $s'=s$. Then, the difference of the two conformal
weights simplifies
\beq
h_{r,s} - h_{r',s} = (r^2-r'^2) {(m+1) \over 4m} + {(r+r')s \over 2} \;
\bmod 1.
\label{diff}
\eeq
For the generic solutions, $r'=r$ or $m-r$, we find $h_{r,s} -
h_{r',s}=0$ or $\half$. For the exceptional ones at $m=6,10,12$ and $30$, we
simply compute the squares modulo $4m$ of the possible values of $r,r'$ and
differences thereof:
\bea
&& m=6 \;:\quad r^2,r'^2 \in \{1,9\} \qquad \Longrightarrow \qquad 
r^2-r'^2 \in \{0,\pm 8\}, \label{m6} \\
&& m=10 \;:\quad r^2,r'^2 \in \{1,9\} \qquad \Longrightarrow \qquad 
r^2-r'^2 \in \{0,\pm 8\}, \\
&& m=12 \;:\quad r^2,r'^2 \in \{1,25\} \qquad \Longrightarrow \qquad 
r^2-r'^2 \in \{0,24\}, \\
&& m=30\;:\quad \cases{r^2,r'^2=1 & \cr \hbox{or } r^2,r'^2=49 &} \qquad
\Longrightarrow \qquad  r^2-r'^2=0.
\label{m30}
\eea
In all these cases, $r$ and $r'$ are always both odd, so that the second term of
(\ref{diff}) vanishes modulo 1, whereas the first term gives various fractions
depending on $m$: thirds for $m=6$, fifths for $m=10$, halves for $m=12$, and
actually integers for $m=30$. Finally, the solutions obtained from the
preceding ones by multiplying $r,r'$ and $m$ by a common integral factor yield
the same fractions, namely no fractions at all, halves, thirds or fifths. 

From (\ref{ttransf}), we therefore arrive at the conclusion that the
coefficients of $M^{(i,ki)}$, for all $i$ and $k$, are integers times second,
third and fifth roots of unity. This implies that, in the sector $\H_{g^i}$,
the charges with respect to the power $g^i$ of the generator are compatible with
the cyclic symmetries made up of $Z_2,Z_3$ and $Z_5$, or equivalently with a
maximal cyclic symmetry equal to $Z_{30}$. What is missing to assert that
this is the maximal cyclic symmetry the whole theory can have, is the value of
the charges with respect to $g$, not $g^i$. In other words, we have to show
that all $M^{(i,1)}$ contain only second, third and fifth roots of unity. 

Let us suppose that there is cyclic symmetry $Z_N$, and let us look at the
matrices $M^{(i,1)} = S^\dagger M^{(-1,i)}S$, for $i=1,2,\ldots,N$. 
From what we have just proved, each matrix $M^{(-1,i)}$ contains only second,
third and fifth roots of unity, and is therefore equal to $M^{(-1,i+30a)}$ for
any integer $a$. Thus $M^{(i,1)} = M^{(i+30a,1)}$. If there exists a
value of $a$ such that $i+30a$ is invertible modulo $N$ (is coprime with $N$),
then $M^{(i+30a,1)}$ hence $M^{(i,1)}$ contains only second, third and fifth
roots of unity (because then $M^{(i+30a,1)}$ is in the set of $M^{(j,kj)}$). 
The only case where there is no such $a$ is when $N$ and $i$ are not coprime
with 30. 

If $i$ is multiple of 5, then $M^{(-1,i)}$ contains only second and third roots
of unity. Indeed $M^{(-1,1)}$ contains second, third and fifth roots of unity,
and yields $M^{(-1,i)}$ upon replacing all roots of unity by their
$i$--th power, effectively setting to 1 all fifth roots of unity. Then
$M^{(-1,i)} = M^{(-1,i+6a)}$ for any $a$, and running through the above argument
with 6 instead of 30 shows that $M^{(i,1)} = M^{(i+6a,1)}$ contains only second,
third and fifth roots of unity except if $N$ and $i$ are not coprime with 6.
Repeating the argument for $i$ a multiple of 3, and then for $i$ a multiple of 
2 eventually proves the statement. 
\bea
\fbox{R4} && \hbox{\it The only cyclic symmetries that unitary minimal conformal
theories can have} \nonumber\\
\noalign{\vspace{-1mm}}
&& \hbox{\it are subgroups of $Z_{30}$.} \nonumber
\eea

\medskip
The above four results are rather general. The next step is to go down into
the details of the various partition functions of Table 1, in order to make use
of their specific form. The easiest are the ``permutation invariants'', namely
those of the diagonal series $(A_{m-1},A_m)$ (for all $m \geq 3$), and of the
complementary series $(A_{m-1},D_{2\ell+3})$ (for $m=4\ell+3$) and
$(D_{2\ell+3},A_m)$ (for $m=4\ell+4$). We begin with these.


\section{Permutation modular invariants}

The modular invariant partition functions we consider in this section have
the form
\beq
Z_{1,1}(\tau) = \sum_{(r,s) \in \KT} \; \chi_{(r,s)}^* \, \chi_{\mu(r,s)}^{},
\eeq
where $\mu$ is a permutation (it is also an automorphism of the fusion rules,
see for instance \cite{dms}). It is the trivial permutation $\mu=1$ for the
diagonal series $(A_{m-1},A_m)$, while for the two complementary series,
\beq
\mu(r,s) = \cases{(r,s) & if $s$ (resp. $r$) is odd, \cr
(m-r,s) & if $s$ (resp. $r$) is even and $r+s \leq m$, \cr
(r,m+1-s) & if $s$ (resp. $r$) is even and $r+s > m$,}
\label{mu}
\eeq
for $m=4\ell+3$ (resp. $m=4\ell+4$). 

For these series, R1 implies that $Z_2$ is the maximal cyclic symmetry, with
four corresponding matrices, $M^{(0,0)}=\delta_{j,\mu(i)}$, $M^{(1,0)},
M^{(0,1)}$ and $M^{(1,1)}$. It remains to see how many realizations of the
$Z_2$ are compatible with the given $M^{(0,0)}$. 

\subsection{The diagonal series}

The matrix $M^{(0,0)}$ is the identity and thus $M^{(0,1)} = \epsilon_i \,
\delta_{i,j}$ is diagonal with signs on the diagonal. It follows that
$M^{(0,1)}$, like $S$, is equal to its own inverse, and the relation
$M^{(1,0)}= S \, M^{(0,1)} \, S$ shows that the same is true of
$M^{(1,0)}$. Since the latter has non--negative integral entries, it must be a
permutation matrix. We set $M^{(1,0)}_{ij} = \delta_{j,\pi(i)}$. 

Because $M^{(1,0)} \, S = S \, M^{(0,1)}$, the permutation $\pi$ must satisfy
\beq
S_{\pi(i),j} = \epsilon_j \, S_{i,j}, \qquad \epsilon_j=\pm 1.
\label{perm}
\eeq
The signs $\epsilon_i$ are the $Z_2$--parities of the states in the periodic
sector. The vacuum being neutral, one has $\epsilon_{(1,1)}=+1$. For
$i=j=(1,1)$, and setting $\pi(1,1)=(r,s)$, the previous equation then says
\beq
S_{\pi(1,1),(1,1)} = S_{(1,1),(1,1)} \qquad  \Longleftrightarrow \qquad 
\sin{\textstyle {\pi r \over m}}\,\sin{\textstyle {\pi s \over m+1}} = 
\sin{\textstyle {\pi \over m}}\,\sin{\textstyle {\pi \over m+1}}.
\eeq
The only integer solutions are $(r,s)=(1,1),(m-1,1),(1,m)$ and $(m-1,m)$ (use
Galois !), but only (1,1) and $(m-1,1)$ are in the Kac table. Moreover
$\pi(1,1)=(1,1)$ must be rejected since $M^{(1,0)}_{(1,1),(1,1)}=1$ would mean
that the vacuum belongs to two distinct sectors. Therefore $\pi(1,1)=(m-1,1)$. 

Inserting $i=(1,1)$ and $j=(r,s)$ in (\ref{perm}) then yields
\beq
S_{(m-1,1),(r,s)} = (-1)^{(m+1)r+ms+1} \, S_{(1,1),(r,s)} = 
\epsilon_{(r,s)} \, S_{(1,1),(r,s)},
\eeq
where the first equality follows from the symmetry of $S$, mentioned in
(\ref{symm}). Since all numbers $S_{(1,1),(r,s)}$ are non--zero, the
previous equation fixes all parities to be $\epsilon_{(r,s)} = 
(-1)^{(m+1)r+ms+1}$. So $M^{(0,1)}$ is uniquely fixed, and yields unique
$M^{(1,0)}$ and $M^{(1,1)}$ by modular transformations.

We have proved our claim: in the models of the diagonal series, there is a
unique way of realizing a $Z_2$ symmetry, which is therefore their maximal
symmetry group. There are only two sectors, which we may call periodic and
antiperiodic, both carrying non--trivial $Z_2$--charges. The field
content and the charges of the two sectors can be read off from the partition
functions, which were found in \cite{zuber}. We reproduce them here
for the sake of completeness,
\bea
&& Z_{PA} = Z_{1,g} = \sum_{(r,s) \in \KT} (-1)^{(m+1)r+ms+1} \,
|\chi_{(r,s)}|^2 ,\\ 
&& Z_{AA} = Z_{g,g} = \sum_{r+s\leq m} 
(-1)^{(m+1)(r+m/2)+ms}\,\chi^*_{(r,s)}\,\chi^{}_{(m-r,s)} \nonumber \\
&& \hskip 4truecm  + \sum_{r+s\geq m+1} 
(-1)^{(m+1)(r+m/2)+ms}\,\chi^*_{(r,s)}\,\chi^{}_{(r,m+1-s)}.
\eea
The other two functions $Z_{PP}$ and $Z_{AP}$ follow from these two by dropping
the signs. 

The first model $m=3$ corresponds to the universality class of the Ising model,
for which one recovers the well--known parity assignments \cite{cardy2}:
in the periodic sector, the identity and the energy density are even while the
spin variable is odd; in the antiperiodic sector, the disorder variable is
even, while the two fermionic degrees of freedom are odd.

\subsection{The complementary series}

We will detail the analysis for one of the two series, say $m=4\ell+3$. The
other, for $m=4\ell+4$, can be treated by the same method. 

For $m=4\ell+3$, the matrix $M^{(0,1)}$ is a generalized permutation matrix,
almost diagonal, 
\arraycolsep=1pt
\beq
M^{(0,1)} = \left(
\begin{array}{ccccccc}
1 & & & & & & \\
 & \pm 1 & & & & & \\
 & & \ddots & & & & \\
 & & & \pm 1 & & & \\
 & & & & \fbox{$\begin{array}{cc} 0 & \pm 1 \\ \pm 1 & 0 \end{array}$} & & \\
 & & & & & \fbox{$\begin{array}{cc} 0 & \pm 1 \\ \pm 1 & 0 \end{array}$} & \\
 & & & & & &  \ddots 
\end{array} \right).
\eeq
The diagonal coefficients refer to the indices $(r,s)$ with $s$ odd, and also
to $(r,2\ell+2)$, whereas the two--by--two blocks are labelled by pairs $(r,s)$
and $(m-r,s)$ or $(r,m+1-s)$ with $s$ even (see above, equation (\ref{mu})). The
first diagonal entry, equal to 1, refers to the index $(1,1)$ ---the vacuum is
neutral---, but otherwise all other signs are uncorrelated. 

Whatever the signs in $M^{(0,1)}$ are, the fourth power of the matrix is
equal to 1, implying the same for $M^{(1,0)}$. Thus $(M^{(1,0)})^{-1} =
(M^{(1,0)})^3$ is a non--negative integral matrix, hence a permutation matrix.
We set again $M^{(1,0)}_{ij} = \delta_{j,\pi(i)}$. 

As before, the relation $M^{(1,0)} \, S = S \, M^{(0,1)}$ implies
\beq
S_{\pi(i),j} = \sum_k \; S^{}_{ik} \, M^{(0,1)}_{kj}.
\label{perm2}
\eeq
Letting $i=j=(1,1)$ yields the same equation we had in the diagonal series, and
the same result $\pi(1,1)=(m-1,1)$. Then for $i=(1,1)$, by using once more the
symmetry (\ref{symm}) of $S$ and the odd character of $m$, one finds from
(\ref{perm2})
\beq
S_{(m-1,1),(r',s')} = (-1)^{s'+1} \, S_{(1,1),(r',s')} = \sum_{k \in \KT} \;
S^{}_{(1,1),k} \, M^{(0,1)}_{k,(r',s')}.
\eeq

This last equation determines all signs in $M^{(0,1)}$ in a unique way. The
calculations are straightforward, so we only quote the result. All diagonal
entries of $M^{(0,1)}$ must be $+1$, except those with indices $(r,2\ell+2)$ for
all $r$, which must be $-1$. All two--by--two blocks must be $\big({0 \atop -1}
\; {-1 \atop 0}\big)$. This can be summarized by saying that
$M^{(0,1)}_{(r,s),(r',s')} = (-1)^{s+1} \, M^{(0,0)}_{(r,s),(r',s')}$. 

Like in the diagonal series, $M^{(0,0)}$ fixes uniquely the other three
matrices, meaning that the $Z_2$ has a unique realization, and is therefore the
maximal symmetry. The same conclusion holds for the other complementary
series, for $m=4\ell+4$. 

We finish by displaying the partition functions. For the series
$(A_{m-1},D_{2\ell+3})$,
\bea
&& Z_{PA} = \sum_{(r,s) \in \KT \atop s \ {\rm odd}} |\chi_{(r,s)}|^2 -
\sum_{(r,s) \in \KT \atop r+s \leq m, \, s \ {\rm even}} \chi^*_{(r,s)} \,
\chi^{}_{(m-r,s)} - \sum_{(r,s) \in \KT \atop r+s > m, \, s \ {\rm even}}
\chi^*_{(r,s)} \, \chi^{}_{(r,m+1-s)}, \qquad \\
&& Z_{AA} = \sum_{(r,s) \in \KT \atop s \ {\rm even}} |\chi_{(r,s)}|^2
- \sum_{(r,s) \in \KT \atop r+s \leq m, \, s \ {\rm odd}} \chi^*_{(r,s)} \,
\chi^{}_{(m-r,s)} - \sum_{(r,s) \in \KT \atop r+s > m, \, s \ {\rm odd}}
\chi^*_{(r,s)} \, \chi^{}_{(r,m+1-s)}, \qquad 
\eea
and for the series $(D_{2\ell+3},A_m)$,
\bea
&& Z_{PA} = \sum_{(r,s) \in \KT \atop r \ {\rm odd}} |\chi_{(r,s)}|^2 -
\sum_{(r,s) \in \KT \atop r+s \leq m, \, r \ {\rm even}} \chi^*_{(r,s)} \,
\chi^{}_{(m-r,s)} - \sum_{(r,s) \in \KT \atop r+s > m, \, r \ {\rm even}}
\chi^*_{(r,s)} \, \chi^{}_{(r,m+1-s)}, \qquad \\
&& Z_{AA} = \sum_{(r,s) \in \KT \atop r \ {\rm even}} |\chi_{(r,s)}|^2
- \sum_{(r,s) \in \KT \atop r+s \leq m, \, r \ {\rm odd}} \chi^*_{(r,s)} \,
\chi^{}_{(m-r,s)} - \sum_{(r,s) \in \KT \atop r+s > m, \, r \ {\rm odd}}
\chi^*_{(r,s)} \, \chi^{}_{(r,m+1-s)}. \qquad 
\eea


\section{More complementary modular invariants}

We analyze in this section the complementary series of partition functions
occurring for $m=4\ell+1$ and $4\ell+2$. These modular invariants are not
permutation invariants, but have a block diagonal form. They can however be
considered as permutation invariants (with the trivial permutation) in terms of
extended characters, written as linear combinations of conformal characters,
and which are in fact the characters of a larger chiral algebra that indeed
extends the Virasoro algebra.

For the problem of symmetries, these two series are more difficult. The basic
reason for this is that some of the coefficients in the sesquilinear forms are
bigger than 1. This means that R1 does not apply, making room 
for richer symmetries. Indeed at least two models are known to have the
permutation group $S_3$ as symmetry, namely the critical and tricritical
3--Potts models, corresponding respectively to $m=5$ ($c={4 \over 5}$) and
$m=6$ ($c={6 \over 7}$). We show below that they are the only two models to
have a larger symmetry group, all others having just a $Z_2$.

Again the two distinct series may be treated by the same methods, so we
restrict ourselves to one of them, which we take to be $(D_{2\ell+2},A_m)$,
$m=4\ell+2$. From Table 1, those partition functions read, in terms of labels in
the Kac table,
\beq
Z_{1,1}(\tau) = \sum_{s \leq r \leq 2\ell-1 \atop r\,{\rm odd}} 
|\chi_{(r,s)} + \chi_{(m-r,s)}|^2 + \sum_{2\ell+2 \leq s \leq r \atop r\,{\rm
odd}}  |\chi_{(r,s)} + \chi_{(r,m+1-s)}|^2 
+ \sum_{s=1}^{2\ell+1} \; 2\,|\chi_{(2\ell+1,s)}|^2. 
\label{compl}
\eeq

R4 says that these models can only have three kinds of cyclic symmetries of
order equal to a prime power: $Z_2$, $Z_3$ and $Z_5$. We examine, for each
symmetry in turn, the possible realizations. 

\subsection{Symmetry of order five}

\smallskip \noindent
An order five symmetry requires $m$ to be a multiple of 10. Indeed, if one
looks at the way R4 was proved, the condition for a $Z_5$ symmetry was that
there should be pairs $(r,s),(r',s')$, allowed by the parity rule, such that 
$h_{r,s}-h_{r',s'}=0 \bmod {1 \over 5}$. From the equations (\ref{diff}) to
(\ref{m30}), this required $m$ to be divisible by 10. So we take $m=0 \bmod 5$.

From the argument that had led to R1, the matrix $M^{(0,1)}$ must be real. We
also know from (\ref{m01}) that a coefficient of $M^{(0,1)}$ is a sum of $n$
fifth roots of unity if the corresponding entry of $M^{(0,0)}$ is equal to
$n$. Here $n=0,1$ and 2, so that $M^{(0,1)}$ is actually equal to $M^{(0,0)}$
except for the diagonal terms of indices $(2\ell+1,s)$. Thus
\beq
M^{(0,1)}_{(r,s),(r',s')} = M^{(0,0)}_{(r,s),(r',s')} + 
\Big[2\cos{2\pi a_s \over 5} - 2\Big] \,
\delta_{r,2\ell+1}\,\delta_{r',2\ell+1}\,
\delta_{s,s'},
\eeq
for some integers $a_s$ between 0 and 4.

R2 says how this matrix must change under Galois transformations. In particular,
it says that for any $\s$ in the Galois group,
\beq
\s\Big(M^{(0,1)}_{(2\ell+1,s),(2\ell+1,s)}\Big) = \s\Big(2\cos{2\pi a_s \over
5}\Big) = M^{(0,1)}_{\s(2\ell+1,s),\s(2\ell+1,s)}.
\eeq

The relevant Galois group is Gal($\Q(\z_{2m(m+1)})$), isomorphic to $\Z^*_{2m}
\times \Z^*_{2(m+1)}$. Let us consider the subgroup $\Z^*_{2m}$ consisting of
those $\s=\s_h$ with $h=1 \bmod 2(m+1)$. Clearly this subgroup is the
Galois group of $\Q(\z_{2m})$.

The way $\s(r,s)$ is computed has been recalled in Section 3, and in this
particular case, one may easily check that $\s_h(2\ell+1,s) = (2\ell+1,s)$.
Thus the numbers $2\cos{2\pi a_s \over 5}$, lying in $\Q(\z_{2m})$ since $m=0
\bmod 5$, must be invariant under Gal($\Q(\z_{2m})$), and so 
must all be rational numbers, equal to 2. It means that all charges in the
periodic sector are equal to zero, and that $M^{(0,i)} = M^{(0,0)}$.

The matrices $M^{(0,i)}$ are therefore both $T$ and $S$--invariant, and this
forces all $M^{(i,j)}$ to be equal to $M^{(0,0)}$. That is, the $Z_5$ can only
be trivially realized.

\subsection{Symmetry of order three}

\smallskip \noindent
For an order three symmetry, $m$ must be a multiple of 6. For the same reason as
in the previous case, the form of $M^{(0,1)}$ is
\beq
M^{(0,1)}_{(r,s),(r',s')} = M^{(0,0)}_{(r,s),(r',s')} + 
\Big[2\cos{2\pi a_s \over 3} - 2\Big] \,
\delta_{r,2\ell+1}\,\delta_{r',2\ell+1}\,
\delta_{s,s'},
\eeq
where $a_s$ are now integers taken modulo 3. We look for a set of nine
matrices $M^{(i,j)}$ consistent with a $Z_3$ symmetry. The integers $i$ and $j$
are taken modulo 3.

Let us first suppose that $M^{(1,1)}$ does not contain third roots of unity.
Then $M^{(2,2)} = M^{(-2,-2)} = M^{(1,1)}$ would not contain
any third root of unity either, so that all charges in the sectors $\H_g$ and
$\H_{g^2}$ are zero. In Section 4, we had then argued that in such
circumstances, no $Z_3$ symmetry would be present at all. Therefore a necessary
condition to have a $Z_3$ symmetry is that $M^{(1,1)}$ contain some third roots
of unity.
 
Third roots of unity in $M^{(1,1)}=T\,M^{(1,0)}\,T^\dagger$ are generated from
$M^{(1,0)}$ through a $T$ transformation, and from (\ref{m6}), this happens only
if some among the following entries
\beq
M^{(1,0)}_{(xm/6,s),(x'm/6,s)} \qquad \hbox{or} \quad 
M^{(1,0)}_{(xm/6,s),(x'm/6,m+1-s)}
\label{xx'}
\eeq
are non--zero, where $x=1,5$ and $x'=3$, or vice versa. The equality
$M^{(1,0)} = S \, M^{(0,1)} \, S$ implies that $M^{(1,0)}$ is symmetric, and
that the coefficients in (\ref{xx'}) are pairwise equal. So, without loss of
generality, we may focus on $M^{(1,0)}_{(xm/6,s),(m/2,s)}$ with $x=1,5$. If
these numbers vanish, for $x=1,5$ and all $s$, there can be no $Z_3$ symmetry. 

They can be computed from the $S$ matrix and $M^{(0,1)}$ given above. One finds,
after a few lines calculation, 
\beq
M^{(1,0)}_{(xm/6,s),(m/2,s)} = {24 \over m(m+1)} \; \sum_{s'=1}^{2\ell+1} \;\;
[1 - \delta_{a_{s'},0}] \, \Big(\sin{\pi ss' \over m+1}\Big)^2.
\eeq
These numbers must be positive, which they are, but also integers. The case
where all charges $a_s$ are non--zero gives an upper bound
\beq
M^{(1,0)}_{(xm/6,s),(m/2,s)} \leq {24 \over m(m+1)} \; \sum_{s'=1}^{2\ell+1}
\;\; \Big(\sin{\pi ss' \over m+1}\Big)^2 = {6 \over m}.
\eeq
One sees that unless $m=6$, these coefficients are to vanish, and the $Z_3$
itself vanishes.

For $m=6$, the upper bound is reached if all charges $a_s$ are different from
zero. This fixes $2\cos{2\pi a_s \over 3}=-1$. It yields a unique $M^{(0,1)}$,
and in turn, a unique set of functions $Z_{g^i,g^j}(\tau)$. 

\subsection{Symmetry of order two}

\smallskip \noindent
The most general form of $M^{(0,1)}$ is more complicated than in the previous
two cases. The partition function $Z_{1,1}$, given in (\ref{compl}),
corresponds to a matrix $M^{(0,0)}$ which has two--by--two blocks $\big({1 \atop
1} \; {1 \atop 1}\big)$ and diagonal entries equal to 2 (in addition to a big
substructure equal to zero, corresponding to rows and columns labelled by
$(r,s)$ with $r$ even). In $M^{(0,1)}$, the blocks become something like 
$\big({\epsilon_1  \atop \eta_2} \; {\eta_1 \atop \epsilon_2}\big)$
with $\epsilon_i,\eta_i=\pm 1$ (independent from block to block), while the
diagonal entries can be equal to $-2$, 0 or 2. 

As a first step, we reduce the degrees of freedom in the signs
$\epsilon_i,\eta_i$ by using again the part of R2 concerned with the Galois
transformation of $M^{(0,1)}$. Let us consider the action of $\s_h$ for
$h=m(m+1)-1 \in \Z^*_{2m(m+1)}$. Since $m$ is even, one obtains for $r$ odd,
that $\tilde r_h = m-r$ and $\tilde s_h = s$. Therefore,
\beq
\cases{r \hbox{ odd} &\cr h=m(m+1)-1&} \qquad \Longrightarrow \qquad
\s_h(r,s) = \cases{(m-r,s) & if $r+s \leq m$,\cr
(r,m+1-s) & if $r+s \geq m+1$.}
\eeq

The blocks $\big({\epsilon_1 \atop \eta_2} \; {\eta_1 \atop \epsilon_2}\big)$
are indexed by the pairs of doublets $(r,s),(m-r,s)$ if $r+s \leq m$, and by
$(r,s),(r,m+1-s)$ if $r+s \geq m+1$. To permute the indices according to
$\s_h$ is to exchange $\epsilon_1 \leftrightarrow \epsilon_2$, $\eta_1
\leftrightarrow \eta_2$ in each block. From the Galois transformation of
$M^{(0,1)}$ stated in R2, this exchange should leave the blocks invariant,
since the Galois group has no action on the entries of $M^{(0,1)}$.
Consequently we find $\epsilon_2 = \epsilon_1$, and $\eta_2=\eta_1$ in each
block.   

In order to get further constraints, we look at the square of $M^{(1,0)}$,
related by an $S$ transformation to the square of $M^{(0,1)}$. The one thing
we know about $M^{(1,0)}$ is that its $(1,1),(1,1)$ coefficient is equal to
zero, because the representation containing the vacuum can only occur in the
periodic sector. We first want to prove that the same entry vanishes in
$(M^{(1,0)})^2$. For this, we use the parity rule R2 as well as R3. As a
preliminary remark, we note that the symmetry (\ref{symm}) of the $S$ matrix
and the fact that $M^{(0,1)}$ is zero for rows and columns labelled by $(r,s)$
with $r$ even, imply that
\beq
M^{(1,0)}_{\mu^a(r,s),\mu^b(r',s')} = M^{(1,0)}_{(r,s),(r',s')}, \qquad a,b=0,1,
\label{musymm}
\eeq
where $\mu$ denotes any one of the two transformations $(r,s) \rightarrow
(m-r,s)$ and $(r,s) \rightarrow (r,m+1-s)$.

R2 says that $M^{(1,0)}_{(1,1),(r,s)}$ is possibly non--zero only for those
$(r,s)$ satisfying $\e_\s(1,1) \e_\s(r,s)$ $= +1$ for all $\s$. Then
(\ref{conf_su2}) together with R3 imply that $(r,s)=(1,1)$ or $(m-1,1)$, except
possibly if $m=6,10$ or 30 (12 does not enter because it is not of the form
$4\ell+2$). For $m=6$, $(r,s)$ can also be equal $(3,1)$ but a non--zero
coefficient at that place would induce third roots of unity in $M^{(1,1)}$,
which are incompatible with a $Z_2$ symmetry. Likewise, if $m=10$, $(r,s)$ can
be equal to $(3,1)$ and $(7,1)$, but then $M^{(1,1)}$ would contain fifth roots
of unity. Finally, for $m=30$, $(r,s)$ can be equal to $(11,1)$ and $(19,1)$,
but now non--zero coefficients at those places produce no root of unity in
$M^{(1,1)}$. Thus $m=30$ is a special case that needs a separate treatment. 

For $m \neq 30$, the rows and columns with labels $(1,1)$ and $(m-1,1)$
form a potentially non--zero two--by--two block. According to
(\ref{musymm}), this block must have four equal elements, actually equal to zero
since the $(1,1),(1,1)$ entry is zero. Therefore we conclude that 
$(M^{(1,0)})^2_{(1,1),(1,1)} = 0$.

For $m=30$, one has instead a symmetric four--by--four block which, by using
(\ref{musymm}), must look like\footnote{By using a Galois transformation, one
can prove that $\beta=0$, but this does not seem to add significant
information for what follows.}
\beq
(M^{(1,0)})_{i,j} = \pmatrix{0 & \alpha & \alpha & 0 \cr
\alpha & \beta & \beta & \alpha \cr
\alpha & \beta & \beta & \alpha \cr
0 & \alpha & \alpha & 0}, \qquad i,j=(1,1),(11,1),(19,1),(29,1).
\eeq
In this case, we find $(M^{(1,0)})^2_{(1,1),(1,1)} = 2\alpha^2$, with $\alpha$
integer.

We can now proceed to compute the number $(M^{(1,0)})^2_{(1,1),(1,1)}$ from the
formula relating the square of $M^{(1,0)}$ to the square of $M^{(0,1)}$. 
The latter contain blocks equal to
\beq
\pmatrix{\epsilon_{(r,s)} & \eta_{(r,s)} \cr
\eta_{(r,s)} & \epsilon_{(r,s)}}^2 = 
\pmatrix{2 & 2\epsilon_{(r,s)} \eta_{(r,s)} \cr
2\epsilon_{(r,s)} \eta_{(r,s)} & 2},
\eeq
and diagonal entries equal to 0 and/or $+4$. In contrast, $(M^{(0,0)})^2$
contains blocks $\big({2 \atop 2} \; {2 \atop 2}\big)$ and diagonal entries
equal to
$+4$. We compute
\bea
&& (M^{(1,0)})^2_{(1,1),(1,1)} = \sum_{i,j \in \KT} S_{(1,1),i} \,
[(M^{(0,0)})^2 + (M^{(0,1)})^2 - (M^{(0,0)})^2]_{ij} \, S_{j,(1,1)} \nonumber\\
&& \hspace{1cm} = (M^{(0,0)})^2_{(1,1),(1,1)} + \sum_{i,j \in \KT} S_{(1,1),i}
\, [(M^{(0,1)})^2 - (M^{(0,0)})^2]_{ij} \, S_{j,(1,1)} \nonumber\\
&& \hspace{1cm} = 2 - 4 \sum_{(r,s) \in \KT \atop r \neq 2\ell+1,\,{\rm odd}}
\delta(\epsilon_{(r,s)} \eta_{(r,s)}+1) \; S_{(1,1),(r,s)}^2 
- 4 \sum_{s=1}^{2\ell+1} \; \delta(M^{(0,1)}_{(2\ell+1,s)(2\ell+1,s)}) \;
S_{(1,1),(2\ell+1,s)}^2.\nonumber \\
\eea

For $m \neq 30$, this number must be equal to zero, implying that all Kronecker
deltas must be equal to 1. This means for $M^{(0,1)}$ that $\eta_{(r,s)} =
-\epsilon_{(r,s)}$ in all blocks, and that the $(2\ell+1,s),(2\ell+1,s)$
diagonal entries are all equal to zero. 

For $m=30$, it can also be equal to 2 (it is manisfestly smaller than 2, and
should equal $2\alpha^2$), and this forces the opposite relations for
$M^{(0,1)}$: $\eta_{(r,s)} = \epsilon_{(r,s)}$ in all blocks, and the
$(2\ell+1,s),(2\ell+1,s)$ diagonal entries are equal to $\pm 2$.

The last piece of argument consists in looking at the diagonal terms in the
part of $M^{(1,0)}$ indexed by pairs $(r,s)$ with $r$ even. This is an easy
calculation which yields, for $r$ even,
\beq
M^{(1,0)}_{(r,s),(r,s)} = \sum_{(r',s') \in \KT \atop r' \neq 2\ell+1,\,{\rm
odd}} \epsilon_{(r',s')} \, S_{(r,s),(r',s')} \,[S_{(r',s'),(r,s)} \pm 
S_{\mu(r',s'),(r,s)}].
\label{final}
\eeq
In this expression, $\mu(r',s') = (m-r',s')$ or $(r',m+1-s')$ depending on
which one is in the Kac table, and the $\pm$ is the sign entering the relation 
$\eta_{(r,s)} = \pm \epsilon_{(r,s)}$ (thus always $-$, except perhaps for
$m=30$).

Because $r$ is even, the symmetry of $S$ says $S_{\mu(r',s'),(r,s)} =
-S_{(r',s'),(r,s)}$, so that\footnote{One may include $r'=2\ell+1$ in the
summation since all $S_{(r,s),(2\ell+1,s')}$ are zero anyway for $r$ even.}
\beq
M^{(1,0)}_{(r,s),(r,s)} = \sum_{(r',s') \in \KT \atop r' \,{\rm
odd}} \epsilon_{(r',s')} \,[1 \pm (-1)] \, S^2_{(r,s),(r',s')} \; \leq \;
2 \sum_{(r',s') \in \KT \atop r' \,{\rm odd}} \; S^2_{(r,s),(r',s')} = 1.
\eeq
Being integers, they are equal to 0 or 1, but the important point is that they
are all simultaneously equal to 0 or 1. Indeed one of them being equal to 1
requires to take the $-$ sign and all $\epsilon_{(r,s)}=+1$.

We finally conclude by showing that $M^{(1,0)}_{(r,s),(r,s)} = 0$ for all $r$
even, implies that there is no $Z_2$ at all. The proof relies once more on the
parity rule. We have argued before that if $M^{(1,1)}$ contains no second roots
of unity (signs !), then there is no $Z_2$. As usually, we relate the signs in
$M^{(1,1)}$ to the $T$ transform of $M^{(1,0)}$:
\beq
M^{(1,1)}_{(r,s),(r',s')} = e^{2i\pi (h_{r,s}-h_{r',s'})} \; 
M^{(1,0)}_{(r,s),(r',s')}.
\eeq
Since $m$ is even, the parity rule implies $M^{(1,0)}_{(r,s),(r',s')} \neq 0$
for $s'=s$ or $m+1-s$. The possibilities for $r'$ being the same as for
$m-r'$, the identity $h_{m-r',m+1-s'}=h_{r',s'}$ allows us to assume $s'=s$, in
which case 
\beq
h_{r,s} - h_{r',s} = (r^2-r'^2) {(m+1) \over 4m} + {(r+r')s \over 2} \;
\bmod 1.
\eeq
For this difference to be equal to $\half$ (and nothing else), the
possible values for $r'$ consistent with the parity rule are just $r'=m-r$ {\it
with $r$ even}. Hence we obtain that the only source of signs in $M^{(1,1)}$ is
in the coefficients $M^{(1,0)}_{(r,s),(m-r,s)}$ and
$M^{(1,0)}_{(r,s),(r,m+1-s)}$ for $r$ even, which themselves are equal to 
$M^{(1,0)}_{(r,s),(r,s)}$ on account of the identities (\ref{musymm}). 

Therefore we can conclude that the vanishing of all the coefficients
$M^{(1,0)}_{(r,s),(r,s)}$ prevents the existence of a non--trivial $Z_2$. The
only alternative is to fix them all equal to 1, which requires taking the $-$
sign in (\ref{final}) (it rules out the exotic possibility at $m=30$) and all
$\epsilon_{(r,s)} = +1$. This fixes completely the matrix $M^{(0,1)}$, and with
it, $M^{(1,0)}$ and $M^{(1,1)}$.

\medskip
This finishes the analysis for one of the series. Settling the other one is
just of matter of repeating the above arguments. Except for $m=5$, which
is singled out in the analysis of the $Z_3$ symmetry, one arrives at exactly the
same conclusion, namely the maximal symmetry is $Z_2$. We summarize by giving
the four partitions functions for both series. 

For the series $(A_{m-1},D_{2\ell+2})$, corresponding to $m=4\ell+1$, they read
\bea
&& Z_{PP}(\tau) = \sum_{s \leq r \leq 2\ell \atop s\,{\rm odd}} 
|\chi_{(r,s)} + \chi_{(m-r,s)}|^2 + \sum_{2\ell+3 \leq s \leq r \atop s\,{\rm
odd}}  |\chi_{(r,s)} + \chi_{(r,m+1-s)}|^2 
+ \sum_{r=2\ell+1}^{4\ell} 2\,|\chi_{(r,2\ell+1)}|^2, \qquad \\
&& Z_{PA}(\tau) = \sum_{s \leq r \leq 2\ell \atop s\,{\rm odd}} 
|\chi_{(r,s)} - \chi_{(m-r,s)}|^2 + \sum_{2\ell+3 \leq s \leq r \atop s\,{\rm
odd}}  |\chi_{(r,s)} - \chi_{(r,m+1-s)}|^2, \\
&& Z_{AP}(\tau) = \sum_{s \leq r \leq 2\ell \atop s\,{\rm even}} 
|\chi_{(r,s)} + \chi_{(m-r,s)}|^2 + \sum_{2\ell+2 \leq s \leq r \atop s\,{\rm
even}}  |\chi_{(r,s)} + \chi_{(r,m+1-s)}|^2, \label{ap} \\
&& Z_{AA}(\tau) = \sum_{s \leq r \leq 2\ell \atop s\,{\rm even}} 
|\chi_{(r,s)} - \chi_{(m-r,s)}|^2 + \sum_{2\ell+2 \leq s \leq r \atop s\,{\rm
even}}  |\chi_{(r,s)} - \chi_{(r,m+1-s)}|^2,
\eea
while for the series $(D_{2\ell+2},A_m)$, corresponding to $m=4\ell+2$,
\bea
&& Z_{PP}(\tau) = \sum_{s \leq r \leq 2\ell-1 \atop r\,{\rm odd}} 
|\chi_{(r,s)} + \chi_{(m-r,s)}|^2 + \sum_{2\ell+2 \leq s \leq r \atop r\,{\rm
odd}}  |\chi_{(r,s)} + \chi_{(r,m+1-s)}|^2 
+ \sum_{s=1}^{2\ell+1} \; 2\,|\chi_{(2\ell+1,s)}|^2, \qquad \quad \\
&& Z_{PA}(\tau) = \sum_{s \leq r \leq 2\ell-1 \atop r\,{\rm odd}} 
|\chi_{(r,s)} - \chi_{(m-r,s)}|^2 + \sum_{2\ell+2 \leq s \leq r \atop r\,{\rm
odd}}  |\chi_{(r,s)} - \chi_{(r,m+1-s)}|^2, \\
&& Z_{AP}(\tau) = \sum_{s \leq r \leq 2\ell \atop r\,{\rm even}} 
|\chi_{(r,s)} + \chi_{(m-r,s)}|^2 + \sum_{2\ell+2 \leq s \leq r \atop r\,{\rm
even}}  |\chi_{(r,s)} + \chi_{(r,m+1-s)}|^2, \\
&& Z_{AA}(\tau) = \sum_{s \leq r \leq 2\ell \atop r\,{\rm even}} 
|\chi_{(r,s)} - \chi_{(m-r,s)}|^2 + \sum_{2\ell+2 \leq s \leq r \atop r\,{\rm
even}}  |\chi_{(r,s)} - \chi_{(r,m+1-s)}|^2.
\eea

As recalled at the beginning of this section, the fully periodic partition
functions, in both series, look like diagonal modular invariants in terms of
extended characters, signalling the presence of an extended symmetry. One may
note that the other partition functions cannot be written in terms of
those, meaning that this extended symmetry is broken by the twisted boundary
conditions. Equivalently, an analysis based on the extended characters rather
than on the conformal ones, would not reveal any symmetry at all.

The above two sets exhaust the frustrated partition functions in those models,
except if $m=5$ and $m=6$, for which other frustrations are possible. 

\subsection{The 3--Potts models}

The previous subsections show that, in addition to a $Z_2$ symmetry, there is
room for a $Z_3$ symmetry when $m=5$ or 6. It is not difficult to see that, in
both cases, the modular invariant partition functions are indeed compatible
with a unique $Z_3$ symmetry. To find the corresponding partition functions
is an easy matter, which we will not detail. However the full symmetry group of
these models is known to be the permutation group $S_3$, and it is 
instructive to see how this conclusion can be reached in the present context. 

For concreteness, we discuss the case $m=5$, describing the
critical point of the 3--Potts model, the other case being exactly similar.
The content and the charges in the various sectors were determined by von Gehlen
and Rittenberg \cite{vonG_R} and by Cardy \cite{cardy2}. An early study of the
3--Potts model using conformal field theoretic techniques was made by Dotsenko
\cite{dotsenko}.

We start by giving the partition functions pertaining to the $Z_2$ and $Z_3$
boundary conditions. For convenience, we indicate, as superscripts, the
conformal weights of the various primary fields. Our notation is that $g$ is
the generator of $Z_2$, and $r$ is a generator of $Z_3$. For a $Z_2$
frustration, they are
\bea
Z_{1,1} &=& |\chi^0_{(1,1)} + \chi^3_{(4,1)}|^2 +
|\chi^{2/5}_{(2,1)} + \chi^{7/5}_{(3,1)}|^2 + 2|\chi^{2/3}_{(4,3)}|^2 +
2|\chi^{1/15}_{(3,3)}|^2,\\
\noalign{\medskip}
Z_{1,g} &=& |\chi^0_{(1,1)} - \chi^3_{(4,1)}|^2 + |\chi^{2/5}_{(2,1)} -
\chi^{7/5}_{(3,1)}|^2,\\
\noalign{\medskip}
Z_{g,g} &=& |\chi^{1/40}_{(2,2)} - \chi^{21/40}_{(3,2)}|^2 + 
|\chi^{13/8}_{(4,2)} - \chi^{1/8}_{(4,4)}|^2.
\eea
Those for a $Z_3$ frustration read (with $\omega=e^{2i\pi/3}$)
\bea
Z_{1,r} &=& Z_{1,r^2} = |\chi^0_{(1,1)} + \chi^3_{(4,1)}|^2 + 
|\chi^{2/5}_{(2,1)} + \chi^{7/5}_{(3,1)}|^2 + (\omega +
\omega^2)|\chi^{2/3}_{(4,3)}|^2 + (\omega + \omega^2)|\chi^{1/15}_{(3,3)}|^2,
\qquad \\
\noalign{\medskip}
Z_{r,r^j} &=& Z_{r^2,r^j} \nonumber \\
\noalign{\medskip}
&=& \omega^{j}\,[\chi^0_{(1,1)} + \chi^3_{(4,1)}]^* \, \chi^{2/3}_{(4,3)}  
+ \omega^{j}\,[\chi^{2/5}_{(2,1)} + \chi^{7/5}_{(3,1)}]^* \,
\chi^{1/15}_{(3,3)} + {\rm c.c.} + |\chi^{2/3}_{(4,3)}|^2 +
|\chi^{1/15}_{(3,3)}|^2.
\eea

As a first step, one may remark that the $Z_2$ and $Z_3$ symmetries are not
compatible, {\it i.e.} the two types of charges cannot be assigned
simultaneously.  The quickest way to see it is to notice that the partition
function $Z_{g,1}$ is a sesquilinear form with coefficients in the set
$\{0,1\}$. Then the same arguments that had led to R1 show that it cannot be
compatible with a $Z_3$ symmetry. It means that the sector of the theory which
is frustrated by the $Z_2$ does not support a diagonal action of $Z_3$ (and
vice versa). So one cannot make sense of partition functions like $Z_{g,r}$ or
$Z_{r,g}$. It also means that the actions of $r$ and $g$ on the various
representations do not commute.

The same conclusion follows by looking at the functions $Z_{1,g}$ and $Z_{1,r}$
giving the $Z_2$ and $Z_3$--charges of the periodic sector. One sees that the
partition function combining the $Z_2$ and $Z_3$--charges into $Z_6$--charges
ought to be
\beq
|\chi_{(1,1)} - \chi_{(4,1)}|^2 + |\chi_{(2,1)} - \chi_{(3,1)}|^2
+ (\omega - \omega^2)|\chi_{(4,3)}|^2 + (\omega - \omega^2)|\chi_{(3,3)}|^2.
\eeq
However, its $S$ transform is not a sesquilinear form with positive
integral coefficients, as it should be. The form of this would--be partition
function was dictated by the assumption that the generators $g$ and $r$ were
diagonal in the same basis. Clearly the failure of this assumption is another
hint of the non--commutativity of $g$ and $r$. 

One of them, say $r$, can still be diagonalized. Then a non--diagonal action of
$g$, compatible with $Z_{1,g}$, is the one that exchanges the two degenerate
representations, 
\beq
(\R_i \otimes \R_i) \quad \stackrel{g}{\longleftrightarrow} \quad (\R_i \otimes
\R_i)', \qquad i=(4,3),(3,3).
\eeq 
So, on each of the two pairs of degenerate representations, $g$ and $r$ act
respectively as $\big({0 \atop 1}\; {1 \atop 0}\big)$ and $\big({\omega \atop
0}\; {0 \atop \omega^2}\big)$, in a non--diagonal way, but consistent with the
above partition functions. These two matrices clearly generate a matrix group
isomorphic to $S_3$.

Thus the periodic sector possesses an $S_3$ symmetry, which can be identified
with the symmetry of the model. The six partition functions are not all
distinct, as one may check that $Z_{1,1}$, $Z_{1,g} = Z_{1,gr} =
Z_{1,gr^2}$ and $Z_{1,r} = Z_{1,r^2}$. The symmetry is smaller in the frustrated 
sectors, being broken down to $Z_2$ or $Z_3$. A quantum chain model for them can
be found in \cite{agr}, along with the appropriate boundary conditions on the
microscopic quantum variables.

Let us finally say a few words about the extended algebra. The periodic
partition function $Z_{1,1}$ is a modular invariant that looks diagonal with
respect to an extended symmetry algebra, which is in this case the $W_3$
algebra \cite{zamo,faza}. The value $c={4 \over 5}$ belongs to its minimal
series, where the algebra has six completely degenerate representations. The
partition function $Z_{1,1}$ is exactly the diagonal combination of all six
characters (two pairs of representations are inequivalent at the $W_3$
level, and have, within each pair, identical characters). In addition, all
partition functions pertaining to the $Z_3$ twisted boundary conditions are
written in terms of the same $W_3$ representations (with charges as in
\cite{faza}), which is not the case for the
$Z_2$ boundary conditions. The generator of $W_3$ of weight 3 is neutral
with respect to the $Z_3$ but is odd under the $Z_2$ (see the sign in $Z_{1,g}$
popping up in front of $\chi_{(4,1)}^3$). The $W_3$ invariance of the critical
3--Potts model is therefore broken in the $Z_2$ twisted sectors. 

The tricritical 3--Potts model, at $m=6$, is similar. The partition functions
for the $Z_2$ boundary conditions have been given above, Section 6.3, and
those relative to the $Z_3$ can be found in \cite{zuber}. We include them here
for completeness
\bea
Z_{1,r} = Z_{1,r^2} &=& |\chi^0_{(1,1)} + \chi^5_{(5,1)}|^2 + 
|\chi^{1/7}_{(5,5)} + \chi^{22/7}_{(5,2)}|^2 + 
|\chi^{5/7}_{(5,4)} + \chi^{12/7}_{(5,3)}|^2 \nonumber\\ 
\noalign{\medskip}
&+& (\omega + \omega^2)|\chi^{4/3}_{(3,1)}|^2 + 
(\omega + \omega^2)|\chi^{10/21}_{(3,2)}|^2 + 
(\omega +\omega^2)|\chi^{1/21}_{(3,3)}|^2, \qquad \\
\noalign{\medskip}
Z_{r,r^j} = Z_{r^2,r^j} &=& \omega^{2j}\,[\chi^0_{(1,1)} +
\chi^5_{(5,1)}]^* \, \chi^{4/3}_{(3,1)}   + \omega^{2j}\,[\chi^{1/7}_{(5,5)} + 
\chi^{22/7}_{(5,2)}]^* \, \chi^{10/21}_{(3,2)} \nonumber\\
\noalign{\medskip}
&+& \omega^{2j}\,[\chi^{5/7}_{(5,4)} + \chi^{12/7}_{(5,3)}]^* 
\,\chi^{1/21}_{(3,3)} + {\rm c.c.} + |\chi^{4/3}_{(3,1)}|^2 
+ |\chi^{10/21}_{(3,2)}|^2 + |\chi^{1/21}_{(3,3)}|^2.
\eea


\section{Exceptional modular invariants}

We end our analysis by examining the six exceptional modular invariant
partition functions, occurring for $m=11,12,17,18,29$ and $30$. From the
result R1 in Section 4, we know that their maximal cyclic symmetry is $Z_2$. We
want to show here that the first two models, related to the simple Lie algebra
$E_6$, are the only ones to possess a non--trivial symmetry, namely a $Z_2$
symmetry. The other four have no symmetry at all. The same arguments
may be applied to them all. For concreteness, we give some details for
$m=11$ only. 

By using the symmetry of the characters, $\chi_{(r,s)} = \chi_{(m-r,m+1-s)}$,
the modular invariant partition function can be written as
\beq
Z_{1,1} = \sum_{r=1 \atop {\rm odd}}^{10} \; |\chi_{(r,1)}+\chi_{(r,7)}|^2 
+ |\chi_{(r,5)}+\chi_{(r,11)}|^2 + |\chi_{(r,4)}+\chi_{(r,8)}|^2.
\eeq
The pairs labelling the characters are not all in the Kac table, but for the
modular transformations, it does not matter, since $S$ also has the symmetry
$S_{(r,s),(r',s')} = S_{(m-r,m+1-s),(r',s')}$. What is important is that all
pairs appearing in $Z_{1,1}$ are different when brought back in the Kac table.

The corresponding matrix $M^{(0,0)}$ can thus be viewed as a direct sum of five
6--by--6 blocks, labelled by $r$, where each block is itself the direct sum of
three 2--by--2 blocks filled up with 1's. The matrix $M^{(0,1)}$,
which specifies the $Z_2$--charges in the periodic sector, has the same form,
with however the 1's replaced by arbitrary and uncorrelated signs. 

As a first step, we use the Galois transformations in order to constrain these
signs. Since the matrix elements of $M^{(0,1)}$ are invariant under the Galois
group, one finds from R2 that 
\beq
M^{(0,1)}_{(r,s),(r',s')} = M^{(0,1)}_{\s(r,s),\s(r',s')}.
\label{galois}
\eeq
The Galois group relevant to the present case is $\Z_{22 \cdot 12}^* =
\Z_{22}^* \times \Z_{24}^*$. On a pair $(r,s)$, the first factor acts on $r$ and
the second factor acts on $s$. For $h \in \Z_{22}^*$, one computes $\langle hr
\rangle_{22}$, {\it i.e.} the product $hr$ taken modulo 22, and one keeps that
number if it is smaller than 11, and otherwise one replaces it with $22 -
\langle hr \rangle_{22}$. Since that algorithm produces the same result for $h$
and $22-h$, it is enough to consider $P\Z_{22}^* \equiv \Z_{22}^*/\{\pm 1\}$.
The same calculations are done with the other factor $P\Z_{24}^*$, with all
congruences taken modulo 24. 

The form of $M^{(0,1)}$ makes the action of the Galois group rather
transparent. The values of $r$, chosen to be odd between 1 and 10, form
precisely the set $P\Z_{22}^*$. Thus the action of that factor simply maps the
first block labelled by $r=1$ onto the other blocks. The constraint
(\ref{galois}) then implies that the five blocks of $M^{(0,1)}$ must be equal,
say to $A$, a 6--by--6 matrix. Thus, 
\beq
M^{(0,1)}_{(r,s),(r',s')} = A_{s,s'} \, \delta_{r,r'}\,, \hskip 2truecm 
\hbox{for $r \in P\Z^*_{22}$}.
\eeq
The matrix $A_{s,s'}$ has indices $s,s'$ in the set $\{1,7,5,11,4,8\}$, in that
order.

The action on the $s$--labels of the other factor $P\Z_{24}^*$ has the effect of
permuting the entries of $A$, which must satisfy $A_{\s_{h'}(s),\s_{h'}(s')} =
A_{s,s'}$ for all $h' \in P\Z^*_{24} = \{1,5,7,11\}$. This leaves in $A$ six
undetermined signs, 
\beq
A_{s,s'} = \pmatrix{\epsilon_1 & \epsilon_2 & 0 & 0 & 0 & 0 \cr
\epsilon_2 & \epsilon_1 & 0 & 0 & 0 & 0 \cr
0 & 0 & \epsilon_1 & \epsilon_2 & 0 & 0 \cr
0 & 0 & \epsilon_2 & \epsilon_1 & 0 & 0 \cr
0 & 0 & 0 & 0 & \eta_1 & \eta_2 \cr
0 & 0 & 0 & 0 & \eta_3 & \eta_4}.
\eeq
At this stage, the Galois constraints are fully satisfied. 

Because $\epsilon_1$ is eventually the $Z_2$--parity of the vacuum, it must be
equal to $\epsilon_1 = +1$. One can determine $\epsilon_2$ by computing 
$M^{(1,0)}_{(3,3),(3,3)}$. It is independent of the signs $\eta_i$, and is equal
to
\bea
M^{(1,0)}_{(3,3),(3,3)} &=& \sum_{(r,s),(r',s')} S_{(3,3),(r,s)} \, 
M^{(0,1)}_{(r,s),(r',s')} \, S_{(r',s'),(3,3)} \nonumber \\
&=& {8 \over 11\cdot 12} \sum_{r \in P\Z^*_{22}} \; \Big(\sin{3\pi r \over
11}\Big)^2 \; \sum_{s,s'} \; A_{s,s'} \, \sin{3\pi s \over 12} \, \sin{3\pi s'
\over 12} = {1-\epsilon_2 \over 3}.
\eea
This should be a non--negative integer, which forces $\epsilon_2 = +1$.

The last step is to compute the coefficient $M^{(1,0)}_{(1,1),(1,1)}$, which
must be 0 if a non--trivial symmetry is present. One finds
\bea
&& M^{(1,0)}_{(1,1),(1,1)} = M^{(0,0)}_{(1,1),(1,1)} + 
\sum_{(r,s),(r',s')} S_{(1,1),(r,s)} \, 
[M^{(0,1)}_{(r,s),(r',s')} -  M^{(0,0)}_{(r,s),(r',s')}]\, S_{(r',s'),(1,1)}
\nonumber \\
&& \hspace{1cm} = 1 + {\textstyle {8 \over 11\cdot 12}} \sum_{r \in P\Z^*_{22}}
\; \Big(\sin{\pi r \over 11}\Big)^2 \; \sum_{s,s'=4,8} \; (A_{s,s'}-1) \,
\sin{\pi s \over 12} \, \sin{\pi s' \over 12} = {4 + \sum_i \eta_i \over 8}.
\qquad
\eea
A non--trivial symmetry requires to set $\eta_i = -1$ for $i=1,2,3,4$. Then all
signs in $M^{(0,1)}$ are uniquely fixed. It remains to check that the partition
functions obtained by modular transformations are well--behaved, which
they are. 

Therefore, the model $(A_{10},E_6)$, $m=11$, has a unique $Z_2$ symmetry. The
charges of the frustrated sectors, periodic and antiperiodic, can be read off
from the two partition functions (for simplicity, we use pairs of labels which
are not necessarily in the Kac table)
\bea
Z_{PA} &=& \sum_{r=1,\;{\rm odd}}^{10} \; |\chi_{(r,1)}+\chi_{(r,7)}|^2 
+ |\chi_{(r,5)}+\chi_{(r,11)}|^2 - |\chi_{(r,4)}+\chi_{(r,8)}|^2, \\
Z_{AA} &=& \sum_{r=1,\;{\rm odd}}^{10} \; |\chi_{(r,4)}+\chi_{(r,8)}|^2
- \Big\{[\chi_{(r,1)}+\chi_{(r,7)}]^* 
\, [\chi_{(r,5)}+\chi_{(r,11)}] + {\rm c.c.}\Big\}.
\eea

The same arguments may be repeated for the twin model $(E_6,A_{12})$, at
$m=12$. One finds a unique $Z_2$ symmetry, and similar partition functions
\bea
Z_{PA} &=& \sum_{s=1,\;{\rm odd}}^{12} \; |\chi_{(1,s)}+\chi_{(7,s)}|^2 
+ |\chi_{(5,s)}+\chi_{(11,s)}|^2 - |\chi_{(4,s)}+\chi_{(8,s)}|^2, \\
Z_{AA} &=& \sum_{s=1,\;{\rm odd}}^{12} \; |\chi_{(4,s)}+\chi_{(8,s)}|^2
- \Big\{[\chi_{(1,s)}+\chi_{(7,s)}]^* 
\, [\chi_{(5,s)}+\chi_{(11,s)}] + {\rm c.c.}\Big\}.
\eea

Among all unitary minimal models, these two at $m=11$ and 12 are the only
ones for which the extended chiral symmetry is preserved in all sectors. 

The analysis of the last four models proceeds the same way. The Galois symmetry
leaves in $M^{(0,1)}$ five arbitrary signs for $m=17,18$, and four arbitrary
signs for $m=29,30$. In each case, one sign is equal to the charge of the
vacuum, and must be equal to $+1$. Then by looking at specific entries of
$M^{(1,0)}$, one finds that they cannot be made non--negative integers unless
all signs are equal to $+1$. It means that all charges are equal to $+1$, so
there is no symmetry at all. This is what should have been expected, since
the Dynkin diagrams of $E_7$ and $E_8$ have no automorphism.
 

\vskip 1truecm
\section*{Acknowledgment}

P.R. would like to thank J.-B. Zuber for useful discussions.


\end{document}